\documentclass[twocolumn,showpacs,pre,superscriptaddress,preprintnumbers,amsmath,amssymb]{revtex4-1}
\usepackage{graphicx,color}
\usepackage{bm}
\usepackage{epsfig}

\graphicspath{{data/}{figures/}}

\begin{document}

\title{Splitting probabilities for Brownian motion with diffusing
  boundaries: Application to polymer translocation}

\author{Alexander K. Hartmann}
\email{a.hartmann@uni-oldenburg.de}
\affiliation{Institut f\"ur Physik, Universit\"at Oldenburg, D-26111
             Oldenburg, Germany}
\author{Satya N. Majumdar}
\email{satyanarayan.majumdar@cnrs.fr}
\author{Alberto Rosso}
\email{alberto.rosso74@gmail.com}
\affiliation{LPTMS, CNRS, Univ.  Paris-Sud,  Universit\'e Paris-Saclay,
  91405 Orsay,  France}

\begin{abstract}
 
We study the translocation of a polymer chain through a nanopore where the chain length 
fluctuates stochastically due to the polymerization-depolymerization processes at the chain 
ends. We map this process to an equivalent representation where the pore performs a 
stochastic random-walk-like process on a line in the presence of two diffusing sinks on 
either side of it with diffusion constants $D_1$ and $D_3$ respectively. The translocation 
process terminates when the pore hits either of the two outer diffusing sinks. In the case 
where the pore motion itself is diffusive with diffusion constant $D_2$, we compute exactly 
the splitting probability that the pore hits the left (right) sink before hitting the right 
(left) sink. We show that the splitting probability in the presence of mobile sinks is 
rather nontrivial compared to the classical case of immobile sinks (the latter corresponds 
to the case when the chain length is fixed). Furthermore, we also compute exactly the 
probability distribution of the translocation time and that of the chain length at the 
completion time of the translocation. We show that both distributions have power law tails 
with exponents that depend continuously on the diffusion constants $D_1$, $D_2$ and $D_3$. 
We validate our analytical predictions via numerical simulations. We then present numerical 
results for the case when the pore performs a fractional Brownian motion with Hurst exponent 
$0<H<1$, while the sinks are still diffusive.

\end{abstract}


\maketitle

\date{\today}

\section{Introduction: translocation of a polymer chain through a pore}

\begin{figure}
\includegraphics[width=0.48\textwidth]{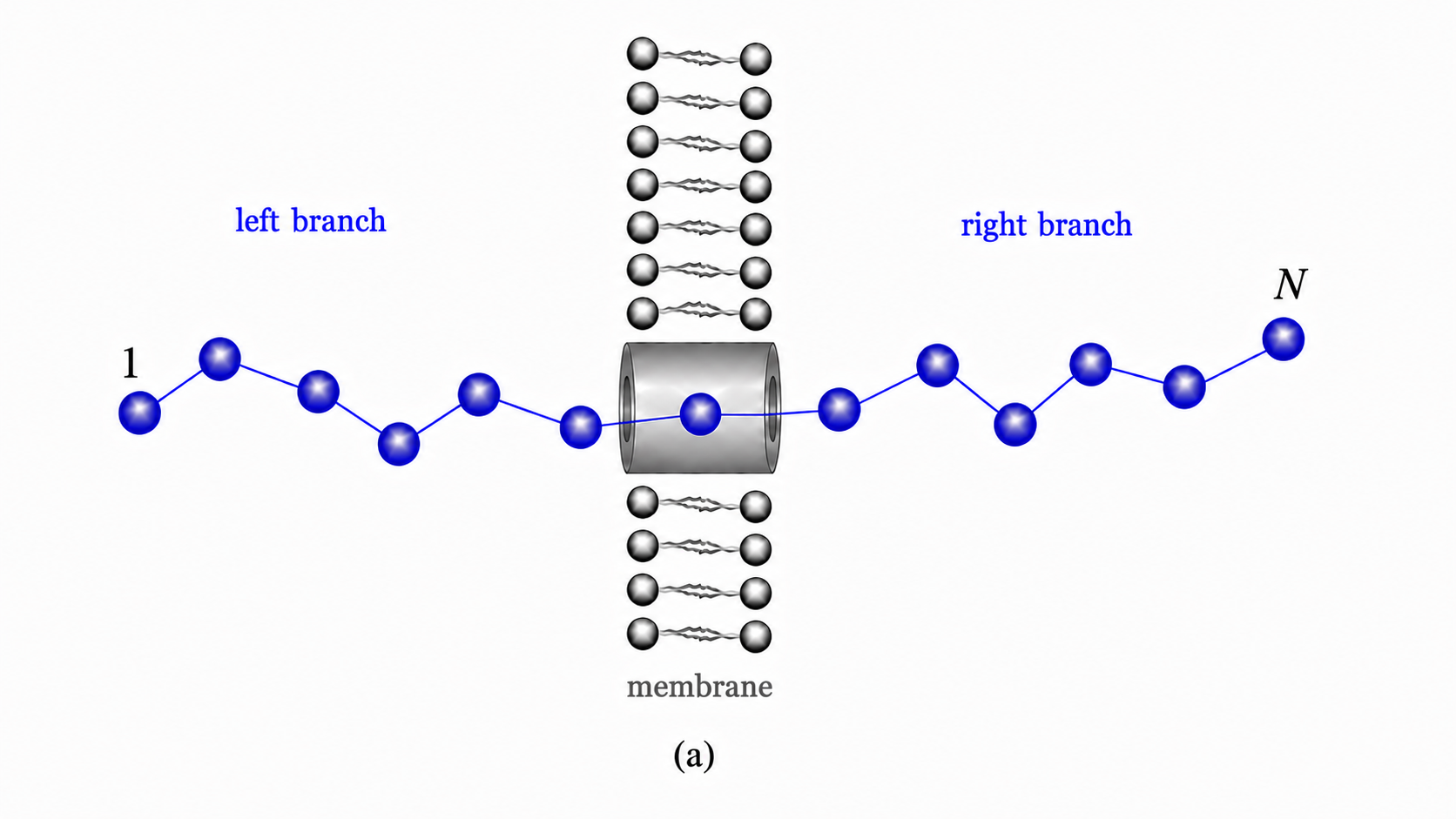}
\includegraphics[width=0.48\textwidth]{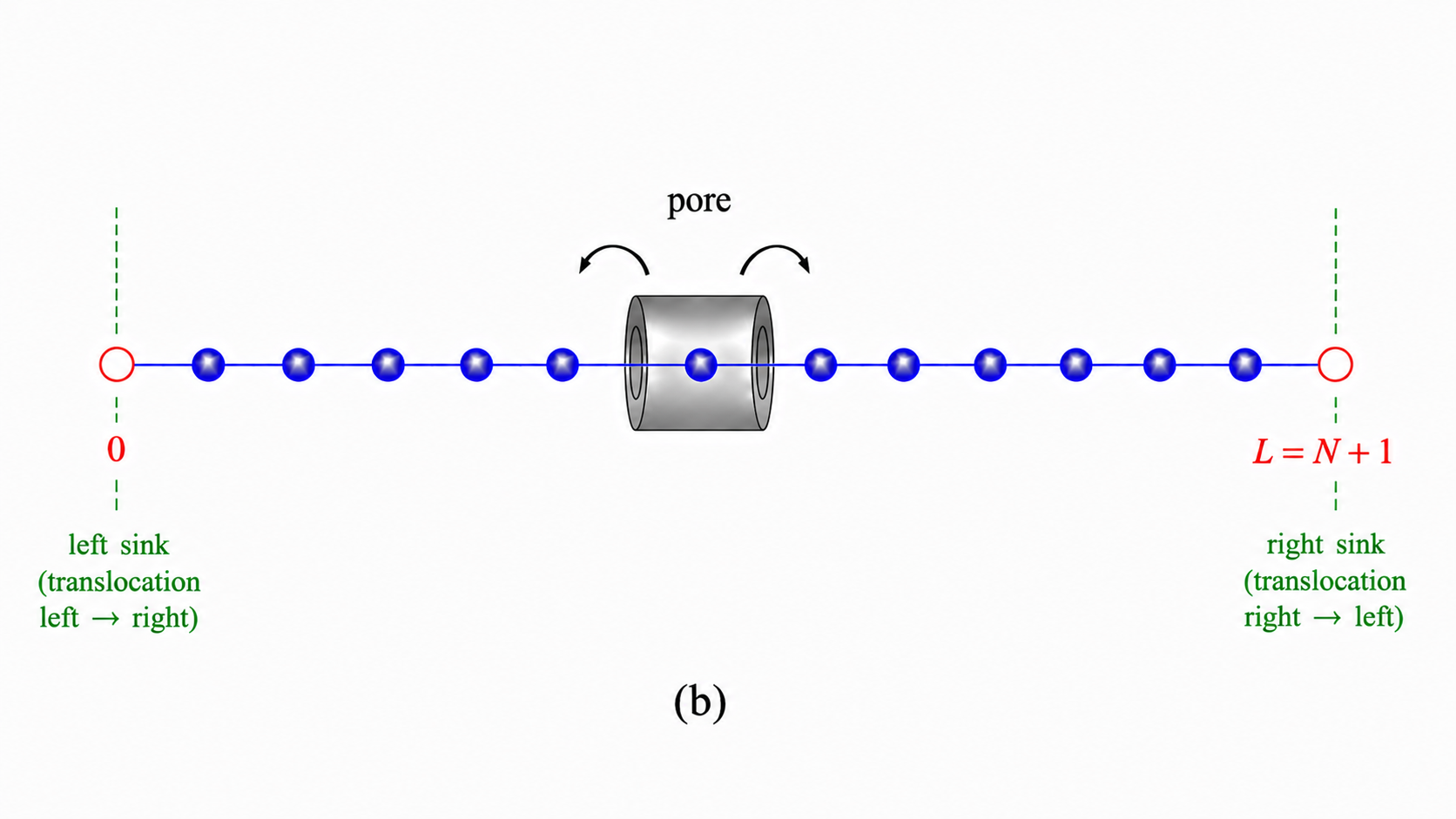}
\caption{ (a) A schematic picture of a polymer chain, consisting of $N$ monomers, translocating
through a pore. The translocation process ends when the number of monomers to the right of the pore
becomes either $N$ (the chain then successfully translocates from left to right)
or $0$ (the chain translocates from right to left). (b) An alternative equivalent representation 
where the pore itself performs a stochastic nearest neighbour hopping on a 
lattice with $L=N+1$ sites (labelled as 
$0,1,2,\ldots, L=N+1$), 
and the process ends when the pore hits either the site $0$ (translocation from left to right of the pore) or 
the site $L$ (translocation from right to left of the pore). 
The two sites $0$ and $L$ (marked by dashed vertical lines)
represent two sink sites for the `moving' pore.}
\label{fig:trans}
\end{figure}

The process by which a polymer chain passes through a narrow pore on a wall/membrane from one side to the other is known as 
the translocation process. Understanding this process is important in many physical, chemical 
and biological systems, and 
it has therefore been studied extensively over the past decades, leading to numerous 
applications\cite{Kasianowicz1996,Meller2001,Meller2003,Grosberg2006,Muthukumar1999,LubenskyNelson1999,Storm2005}. In Fig. 
\ref{fig:trans} (a), we show a schematic picture of this process for a polymer chain consisting 
of $N$ monomers. 
The number of monomers, $s(t)$, to the right of the pore at time t, usually referred to 
as the translocation coordinate, is a key quantity for understanding the translocation 
process~\cite{Muthukumar1999,LubenskyNelson1999,Storm2005}.
As a monomer 
passes through the pore to its right (left), the translocation coordinate $s(t)$ increases (decreases) by $1$. Thus the 
translocation coordinate $s(t)$ undergoes a random-walk-like stochastic process~\cite{LubenskyNelson1999,Chuang2001,KantorKardar2004,Chatelin2008,Zoia2009}. The translocation process becomes complete 
when $s(t)$ either hits $N$ or $0$ for the first time. In the former case when $s(t)$ hits $N$ before hitting $0$, the chain 
translocates successfully from the left to the right of the pore. In contrast, if $s(t)$ hits $0$ before hitting $N$, then the 
chain translocates from the right to the left of the pore. An alternative, but completely equivalent, description 
is to consider the 
pore itself as performing a stochastic hopping process on a fixed
one-dimensional lattice  whose sites are labelled 
$i=0,1,2,\ldots, N, N+1=L$, see Fig.~\ref{fig:trans}~(a).
The two sites $0$ and $L$ at the two ends act like sinks. Whenever 
the `stochastic' pore reaches either $0$ or $L$, the process stops. If it reaches $0$ first, this means that the number of 
monomers $s(t)$ to the right of the pore hits $N$ and the translocation occurs from the left to the right of the pore. In 
contrast, if the stochastic pore reaches $L=N+1$ first, then the number of monomers to its right is $0$ indicating translocation 
from the right to the left of the pore. We will refer to this alternative picture as representation (b) in the rest of the paper. 
We will see later that this representation (b) of the translocation process is convenient for the problem addressed in this 
paper.

Several studies have shown that
the actual stochastic motion of the translocation coordinate $s(t)$, or equivalently that of the motion
of the pore in representation (b) mentioned above,
can be very complex as it depends on various factors such as the pore size, the solvent concentration, the
excluded volume effect etc.~\cite{LubenskyNelson1999,Chuang2001,Chatelin2008,Zoia2009}.
For large $N$, one can make a further approximation by replacing the lattice
by a continuous interval over $[0,L]$ and assuming the translocation coordinate
$s(t)$ is a continuous stochastic process in time.

\begin{figure}
  \begin{center}
\includegraphics[width=0.5\textwidth]{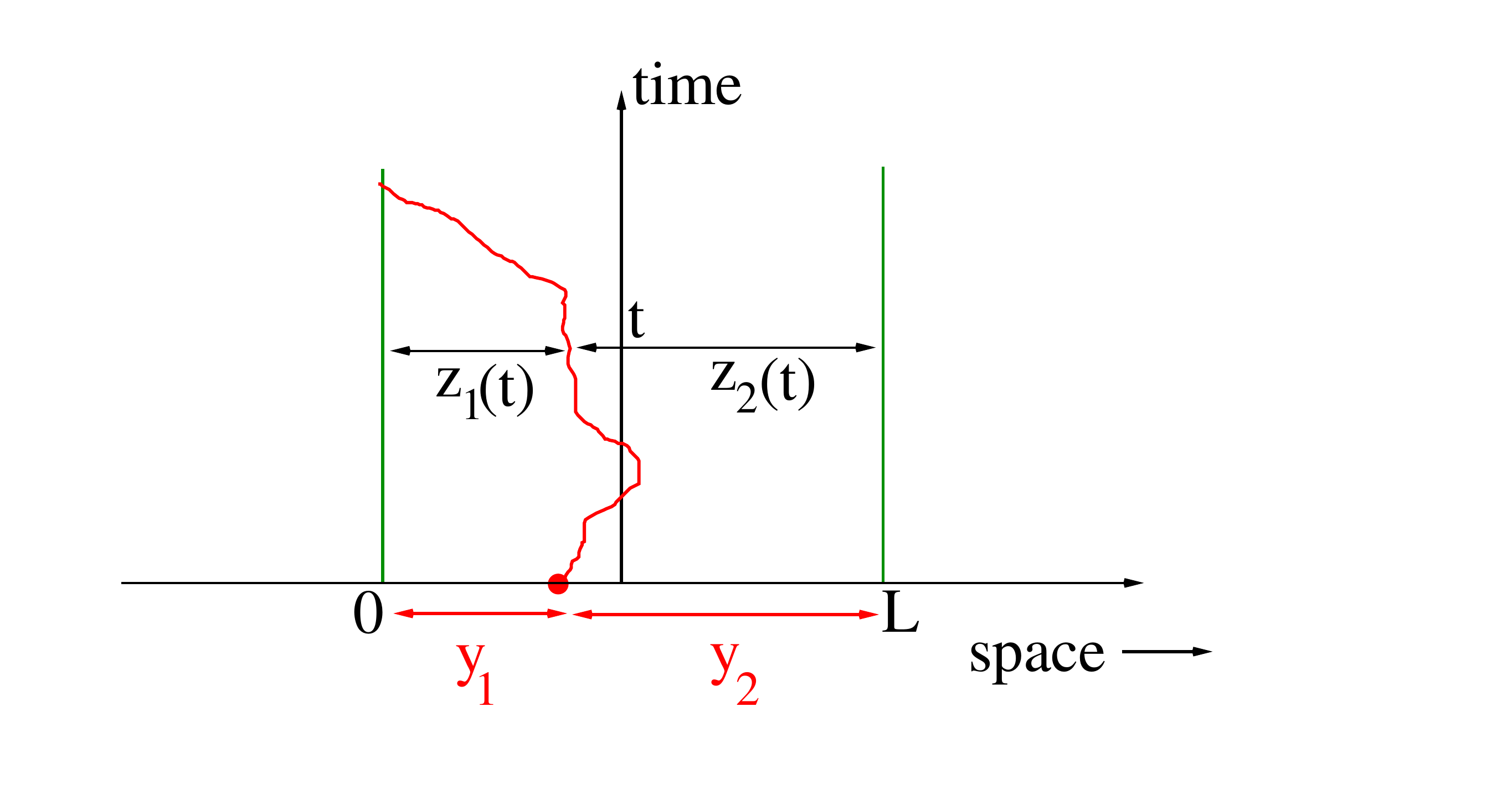}
  \end{center}
\caption{A schematic trajectory of a single Brownian particle (shown by red solid line)
diffusing in a box $[0,L]$, starting initially at $0\le y_1\le L$. The process
stops when the walker hits either of the two boundaries at $0$ or at $L$. In the displayed configuration,
the walker hits the left boundary at $0$ before hitting the right boundary.}
\label{fig:1bm}
\end{figure}

Although the actual stochastic dynamics of the pore may be rather complicated in reality,
it has been useful to study some generic properties of the translocation process via
modeling it by some well-known stochastic process, such as the
fractional Brownian motion (fBM) with Hurst exponent $0<H<1$~\cite{Zoia2009}.
An fBM is a Gaussian process with zero mean and a correlator~\cite{fBM1968}
\begin{equation}
\langle s(t_1)s(t_2)\rangle= D\, \left[t_1^{2H}+t_2^{2H}- |t_1-t_2|^{2H}\right]\, ,
\label{fBM_def}
\end{equation}
where $D$ is a non-negative constant.
Note that for general $0<H<1$, the fBM is a non-Markovian process, except for $H=1/2$, where
it reduces the ordinary Brownian motion which is a Markov process~\cite{Krug1997,WMR2011}.
In this paper, we will first focus
on ordinary Brownian motion $(H=1/2)$ for which several results can be derived analytically and
then present some numerical results for other values of $H$.
In representation (b) and in the continuum, the case $H=1/2$ corresponds
to the pore performing a Brownian motion with diffusion constant $D$ in an interval  $x\in [0, L]$ with
two absorbing walls at $x=0$ and $x=L$, see Fig.~\ref{fig:1bm}.
The location of the pore at time $t$
is denoted by $z_1(t)$ and hence, in this picture, the translocation coordinate is simply
$s(t)= z_2(t)=L-z_1(t)$, as shown in Fig. \ref{fig:1bm}. The pore starts from the initial location $y_1$, i.e.,
the initial value of the translocation coordinate is $y_2=L-y_1$. The process stops when the Brownian pore
hits either the wall at $0$ (indicating translocation of the polymer chain from the left to the
right of the pore)
or the wall at $L$ (translocation from right to left).

One can then study various observables associated with the actual translocation process within this
simplified Brownian motion picture, where analytical results can be obtained. For example,
one natural question is: what is the probability that the polymer chain will translocate from the
left to the right of the pore? In the Brownian motion picture of representation (b), this is equivalent to
asking the probability that the Brownian walker in Fig. \ref{fig:1bm}, starting initially at $y_1$,
will hit the left wall at $0$ first before hitting the right wall at $L$. This is the classical
`splitting' probability
(or sometimes called `hitting' probability in the probability literature)~\cite{Feller_book,Redner_book}
of a Brownian motion starting at $y_1=L-y_2$. In the context of the
translocation process, $y_2$ is  the initial value
of the translocation coordinate $z_2(t=0)$. Expressing everything
in terms of the initial translocation coordinate $y_2$,
for the walker, starting at $y_1=L-y_2$, let
\begin{eqnarray}
  p_{\rm left}(y_2|L) & = &
  {\rm Prob}[ {\rm walker  \, hits\, the\, left\, wall\, first}] \nonumber \\
  p_{\rm right}(y_2|L) & = &
  {\rm Prob}[ {\rm walker\,  hits\, the\, right\, wall\, first}]  
\label{def:p:left:right}
\end{eqnarray}
Essentially, the net probability flux   out of the box $[0,L]$
gets split between the left and the right with these probabilities, leading to
the name splitting probabilities.
Evidently, $p_{\rm left}(y_2|L)+p_{\rm right}(y_2|L)=1$. The exact expressions for these splitting
probabilities are incredibly simple~\cite{Feller_book,Redner_book}
\begin{equation}
p_{\rm left}(y_2|L)= \frac{y_2}{L}\,  \quad {\rm and} \quad p_{\rm right}(y_2|L)= 
1- \frac{y_2}{L}\, .
\label{splitting_BM.1}
\end{equation}
Interestingly, the splitting probabilities are independent of the diffusion constant $D$ of the Brownian motion.
A simple derivation of this classical result using the backward Fokker-Planck equation is reproduced
in Section \ref{one:walker} for the sake of completeness, as well as to illustrate
the backward Fokker-Planck method. Splitting probability for several non-Brownian processes
in one dimension has been computed exactly. This includes
L\'evy flights~\cite{Widom1961,Blumenthal1961}, several
anomalous subdiffusive processes~\cite{MRZ2010},
Brownian particle in the presence of an external potential~\cite{MRZ2010},
symmetric jump processes~\cite{Klinger2022}, active run-and-tumble particle
in one dimension~\cite{GT2024,Basu2026,Frydel2026} etc.
For a recent application of splitting probability of $N$ Brownian
motions subjected to simultaneous stochastic resetting triggered by the
crossing of a threshold by any one of
the walkers, see Ref.~\cite{BMP2025}.

\begin{figure}[htb]
\includegraphics[width=0.48\textwidth]{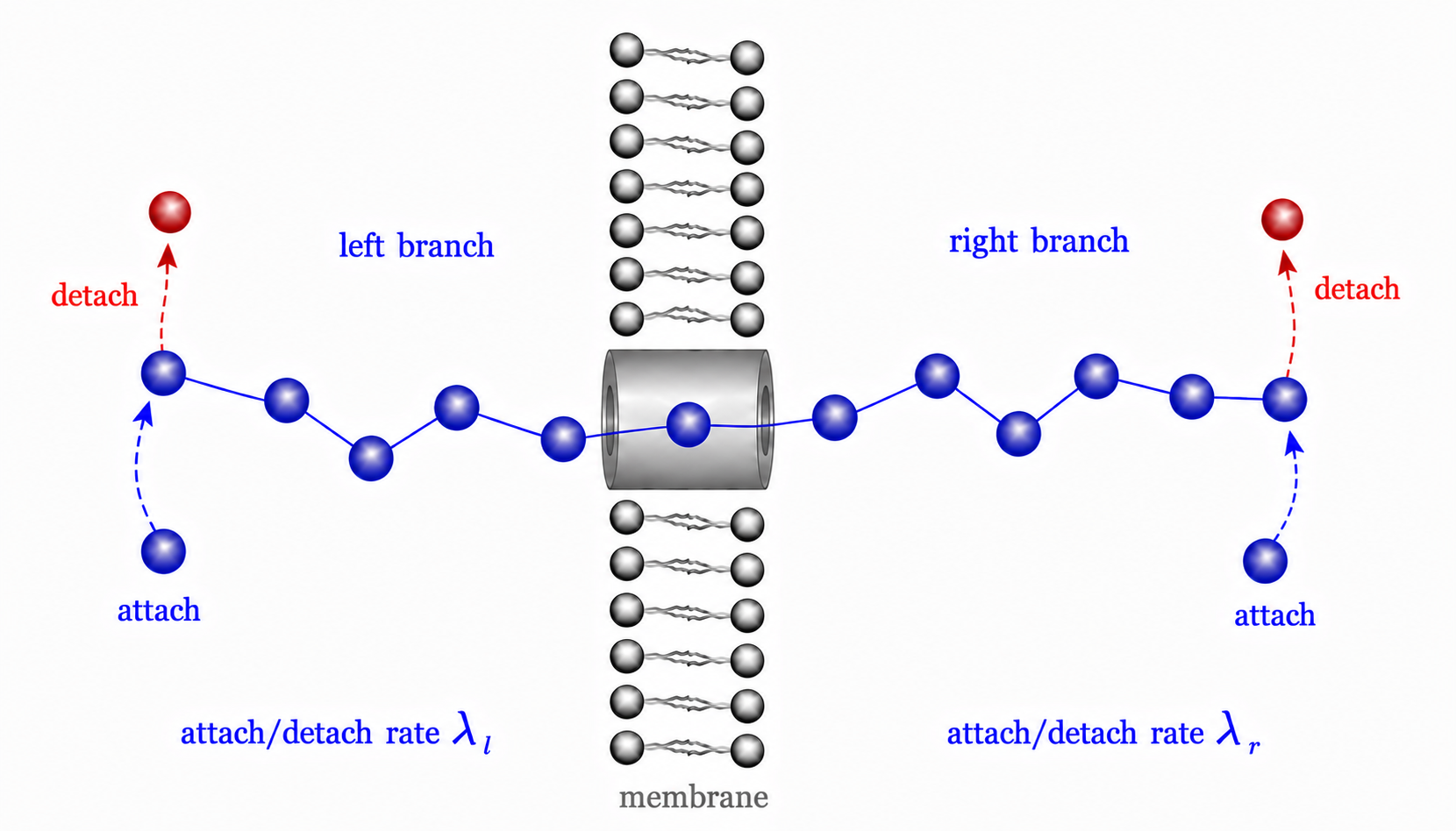}
\includegraphics[width=0.48\textwidth]{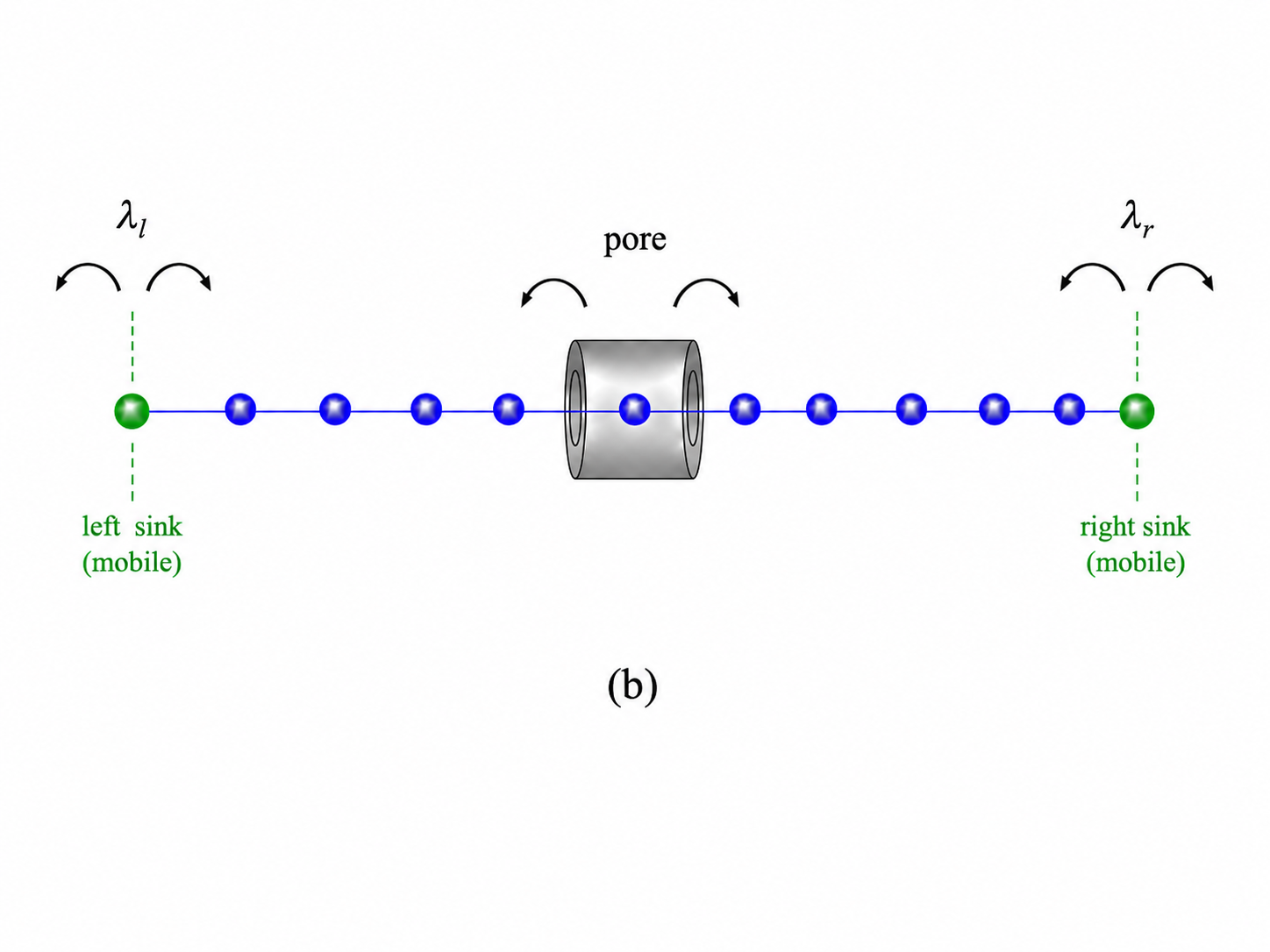}
\caption{ (a) A schematic picture of a polymer chain translocating
through a pore. The monomer located at the left end (shown by
a red filled circle) can detach or attach with rate $\lambda_l$.
Similarly the monomer located at the right edge (also shown by a red filled circle)
can detach or attach with rate $\lambda_r$. 
(b) An alternative equivalent representation 
where the pore itself performs a stochastic nearest neighbour hopping on a lattice.
Also, the two boundaries of the cluster representing the end monomers also
perform `random walk' like hopping to the left and to the right. The two
sites at the end (now also moving stochastically) act like sinks. The translocation process ends when
the central walker hits either the left or the right sink.}
\label{fig:trans2}
\end{figure}

In this paper, we introduce a generalised model of Brownian translocation process where the
chain length $N$ of the polymer is not fixed, but fluctuates in time. For a polymer in a solvent,
often the chain length decreases through depolymerization whereby a single monomer from
either of the two edges can detach from the polymer with
some rates~\cite{deGennes1979,Odian2004,Wang2023,RMFM2025}. Conversely, polymerization may also occur,
whereby a new monomer can attach at either end of the
chain with certain rates. For simplicity, we assume that at each end, the
rates of polymerization and depolymerization are equal, but
these rates can be different at the two ends. We denote these rates by $\lambda_l$
(for the left end) and $\lambda_r$ (for the right end),
see Fig.~\ref{fig:trans2} for a schematic representation.
In the random walk representation (b), this means that the absorbing walls
at $0$ and $L$ are not fixed in time, but their locations also diffuse, in addition to the
Brownian motion of the pore in the middle. Since the left end polymerizes and depolymerizes with
the same rate $\lambda_l$, in the representation (b) and in the continuum Brownian picture, this
corresponds to the case where the left wall, initially located at $0$, diffuses with a certain 
diffusion constant $D_1$
proportional to $\lambda_l$. Similarly, the right wall, initially located at $L$, also diffuses, 
but with a different
diffusion constant $D_3$ proportional to $\lambda_r$. We also assume that the central
Brownian pore, sandwiched between the two diffusing walls, itself diffuses with a diffusion
constant $D_2$. Thus this translocation process for a polymer chain with fluctuating size, via
polymerization/depolymerzation at the edges, then maps directly onto a model of three
`vicious' Brownian walkers on a line with diffusion constants $\{D_1, D_2, D_3\}$ and
starting from initial separations $y_1$ (between the first and the second walker) and
$y_2=L-y_1$ (between the 2nd and the 3rd walker), see Fig.~\ref{fig:trans2}.

These walkers are called `vicious'
because the process terminates whenever any two of them meet meet.
The vicious walker problem with $M$ walkers was introduced by Karlin
and McGregor in the probability literature~\cite{KM1959}
and by de Gennes~\cite{deGennes1968} in the physics literarture, although
the name `vicious' was given by M. E. Fisher in the
context of the dynamics of domain walls in commensurate-incommensurate
system~\cite{Fisher1984},
Since then, the vicious walker problem has been studied extensively, both in the physics and
in the mathematics literature~\cite{HF1984,EG1995,KGV2000,Johansson2002,KT2002,BW2004,RE2005,TW2007,SMCR2008,NM2009,
BFPSW2009,FMS2011,SMCF2013,KMS2014,NR2017,GLMS2019,GMS2021}.
The general $M$ particle problem has mostly been studied in the case when all
the diffusion constants are identical. In that case, conditioned on the process
being alive up to time $t$ (i.e., none of the $M$ walkers meet another member up to $t$),
the joint distribution of the positions of the walkers at time $t$, rescaled by $\sqrt{t}$,
happens to be identical to the joint distribution of $M$ eigenvalues of a real symmetric
$(M\times M)$ Gaussian matrix (the so called Gaussian Orthogonal Ensemble)~\cite{Johansson2002,TW2007,SMCR2008}. However,
for different diffusion constants, the general $M$-particle problem is hard to solve.
For $M=3$ and with different diffusions constants $\{D_1, D_2, D_3\}$ which is 
the case of interest here,
it turns out that several observables can be computed exactly, via a mapping
to a diffusion process in a wedge in two dimensions~\cite{FG1988,BJMKS2003,MB2010}. This includes, in
particular, the survival probability or the persistence of the process up to time $t$~\cite{Redner_book,BMS2013}.
However, we are interested here in the probability of translocation, i.e., the splitting
probabilities $p_{\rm left}(y_2|L)$ and $p_{\rm right}(y_2|L)=1- p_{\rm left}(y_2|L)$ as defined in \eqref{def:p:left:right}, but now for the
case where the walls are Brownian walkers initially seperated at distance $L$.

To the best of our knowledge, the splitting probabilities for three Brownian walkers with
different diffusion constants $\{D_1, D_2, D_3\}$ have not been studied before. In this paper,
we present an exact solution of these splitting probabilities in Section
\ref{Ana_section} (see Eqs. (\ref{final_sol.1})).

In this translocation process of a polymer chain with fluctuating size, there are additional
observables of interest which are completely trivial in the fixed chain problem. For example,
it is natural to ask: what is the distribution of the length of the polymer chain when
it finishes the translocation process? 
In the case of fixed chain length, the length of the polymer chain at the time of the completion
of the translocation process is trivially $N$. However, this distribution turns out to be nontrivial 
in the case of fluctuating chain size, as we show in this paper. 
More generally, one can ask what is the joint probability distribution $P^{\rm left}(t,z)$ 
that the translocation process (say from the left of the pore to its right)
completes at time $t$ and that the length of the chain at the time of the the completion is equal to $z$.
In this paper, for the Brownian version of the translocation process, we compute this
joint distribution $P^{\rm left}(t,z)$ exactly 
in Section \ref{JPDF_tz} (see Eqs. (\ref{scaling.1}) and (\ref{Hu_def})). 
By integrating over $z$ or over $t$, we then compute the marginal distribution of the time of 
completion, as well as the
marginal distribution of the size of the polymer chain at the time of the completion of 
translocation. We show that the marginal distribution of the translocation time $t_{\ell}$ has
a power-law decay for large $t$
\begin{equation}
P^{\rm left}_{t_\ell}(t) = \int_0^{\infty} P^{\rm left}(t, z)\, dz \approx \frac{C_1}{t^{\kappa}}
\label{margt_asymp_intro}
\end{equation}
where the amplitude $C_1$ is a constant that depends on the initial separations 
and the exponent $\kappa$ is given by 
\begin{equation}
\kappa= 1+ \frac{\pi}{2\, \cos^{-1}\left(\frac{D_2}{\sqrt{(D_1+D_2)(D_2+D_3)}}\right)}\, .
\label{kappa_def_intro}
\end{equation}
Thus this exponent depends continuously on the three diffusion constants $D_1$, $D_2$ and $D_3$.
Similarly, we show (see Section \ref{JPDF_tz}) that
the marginal size distribution denoted by $P^{\rm left}_z(z)$ decays as a power law for large $z$
\begin{equation}
P^{\rm left}_z(z)=\int_0^{\infty} P^{\rm left}(t, z)\, dt\approx \frac{C_2}{z^{\beta}} \quad {\rm as}
\quad z\to \infty\, , 
\label{margz_asymp_intro}
\end{equation}
where $C_2$ is a constant (function of the initial separations) and the 
exponent $\beta$ also depends continuously on the diffusion
constants $\{D_1,D_2, D_3\}$. The exact expression for the exponent $\beta$ reads
\begin{equation}
\beta=1+\frac{\pi}{\cos^{-1}\left(\frac{D_2}{\sqrt{(D_1+D_2)(D_2+D_3)}}\right)}\,  .
\label{beta_def_intro}
\end{equation}
These power-law decays of the two marginal distributions in 
Eqs. (\ref{margt_asymp_intro}) and (\ref{margz_asymp_intro})
are two main exact results of this paper.
We also perform extensive numerical simulations to verify our analytical predictions
for the splitting probabilities, the joint distribution $P^{\rm left}(t,z)$ and
the result in Eqs. (\ref{margz_asymp_intro})-(\ref{beta_def_intro}). We find
excellent agreement between the theory and the simulations. 
We also present
numerical results for
the splitting probabilities and the marginal distributions of the translocation time $t_\ell$
and the chain length $z$ at the completion time of the translocation for the case when the central
particle performs an fBM with Hurst exponent $H\ne 1/2$, while the left and the
right particles (the sink sites) perform ordinary Brownian motions.

The rest of the paper is organized as follows. In Section \ref{one:walker}, we first
present an illustration of the backward Fokker-Planck method to derive the well-known
result for the splitting probabilities of a Brownian motion in a fixed interval $[0,L]$.
In Section \ref{3_walkers}, we compute the splitting probability
for a Brownian walker with diffusion constant $D_2$ and surrounded by two sinks performing independent Brownian motions with
diffusion constants $D_1$ and $D_3$.
We first present the numerical approach in Section \ref{Num_section}
and then in Section \ref{Ana_section} we present the exact analytical results.
In Section \ref{JPDF_tz}, we compute exactly, for the three Brownian walkers model, 
the joint distribution of the hitting
time of the central particle to its left neighbour (before the right neighbour)
and the separation from the right neighbour at the time of hitting the left neighbour.  
In Section \ref{fBM_section}, we present numerical results for the case when
the central particle performs an fBM with Hurst exponent $H\ne 1/2$, while
the two outer particles perform ordinary diffusions. Finally, we conclude with
a summary and outlook in Section \ref{summary}.

\section{Splitting probability for a Brownian walker between two fixed walls
\label{one:walker}}

We first recall the well-known results for the splitting probability
of a single Brownian walker diffusing inside an interval $[0,L]$ of length $L$. 
Here the walls/boundaries at $x=0$ and $x=L$ are fixed and do not change
with time. The position $x(t)$ 
of the walker inside the interval evolves by the Langevin equation
\begin{equation}
\frac{dx}{dt}= \sqrt{2D}\, \eta(t)\, ,
\label{bm1.1}
\end{equation}
where $D$ is the diffusion constant and $\eta(t)$ is a zero mean Gaussian white noise
with correlator $\langle \eta(t)\eta(t')\rangle=\delta(t-t')$. The walker starts
from the initial position $0\le x(0)=y_1\le L$ and the process stops
when the walker hits either the left boundary at 0 or the right boundary at $L$.
Let $y_2=L-y_1$ denote its initial
distance from the right wall at $L$ (see Fig.~\ref{fig:1bm}).
Let $z_1(t)$ and $z_2(t)$ denote the distances of the walker from the two walls at time $t$
before it hits either of the walls (see Fig.~\ref{fig:1bm}).
In the context of the translocation process discussed in the introduction, the separation
$z_2(t)$ is precisely the translocation coordinate of the underlying polymer chain.
Hence, $y_2$ is the initial value of the translocation coordinate.

When considering the splitting probabilities
$p_{\rm left}(y_2|L)$ and $p_{\rm right}(y_2|L)=1- p_{\rm left}(y_2|L)$ as defined
in \eqref{def:p:left:right}, it is a classical result in
probability theory that $p_{\rm left}(y_2|L)$ is given by the simple
linear formula~\cite{Feller_book,Redner_book} as shown in
\eqref{splitting_BM.1}.
Note that the splitting probability is independent of the diffusion constant $D$ of the walker.

Equation \eqref{splitting_BM.1} can be derived straightforwardly using the backward Fokker–Planck approach (for different
applications of the backward method, see Refs.~\cite{M2005,BMS2013}). 
Within this approach, it is easier to consider the dependency of the splitting
on the initial position $y_1$, i.e.,
one monitors how the splitting probability $p_{\rm left}(y_1) \equiv
p_{\rm left}(y_2=L-y_1|L) $
changes as one varies the initial position $y_1$.
Here, we do not need to keep track of
the distance $z_1(t)=L-z_2(t)$ of the walker as a function of time $t$. 
Instead, we integrate over the entire history of the process and treat
only the initial position $y_1$ 
of the walker as the relevant variable.
Starting from $y_1$,
during the first interval $\Delta t$, the
walker moves to a new position $y_1+ \sqrt{2D} \eta(0)\, \Delta t$ after
which it continues to diffuse from this new position.
Here, $\eta(0)$ refers to the
first kick (noise). Consequently
\begin{equation}
  p_{\rm left}(y_1)=
  \langle p_{\rm left}(y_1+ \sqrt{2 D} \eta(0)\, \Delta t\,)\rangle\, ,
\label{bm1.3}
\end{equation}
where $\langle \rangle$ refers to the average over the initial noise $\eta(0)$. Expanding
in a Taylor series for small $\Delta t$ up to the second order and using the
properties of the white noise, namely $\langle \eta(0)\rangle=0$ and
$\langle \eta^2(0)\rangle=\frac{1}{\Delta t}$ (which follows from the delta correlator
of the noise), one obtains from Eq.~(\ref{bm1.3}) an ordinary differential equation
\begin{equation}
D\, \frac{d^2 p_{\rm left}(y_1)}{dy_1^2}=0\, .
\label{bm1.4}
\end{equation}
This equation holds in $y_1\in [0,L]$ with the boundary conditions: (i) $p_{\rm left}(y_1=0)=1$
and (ii) $p_{\rm left}(y_1=L)=0$. The boundary condition (i) simply follows from the fact that
if the walker starts at the left boundary, it will immediately exit through the 
left boundary and hence the probability of this event is $1$. Similarly, (ii) follows
from the fact that if the walker starts at the right boundary,
the probability that it will hit the left boundary before hitting the right boundary 
is exactly zero. The solution to this differential equation satisfying the two boundary 
conditions is precisely given by
\begin{eqnarray}
  p_{\rm left}(y_1) & = & 1-\frac{y_1}{L} ,
  \quad {\rm and}\,\, {\rm consequently} \nonumber \\
 p_{\rm right}(y_1) & = &
1- p_{\rm left}{y_1}=\frac{y_1}{L}\, .
\end{eqnarray}
 It is also clear from Eq.~(\ref{bm1.4})
that the diffusion constant $D$ just drops out of the solution.

In the context of the translocation process discussed in the introduction, it is useful
to express these results 
in terms of the initial value of
the translocation coordinate $y_2=L-y_1$, which then leads to
\eqref{splitting_BM.1}.

\section{Splitting probability for a Brownian walker between two
  diffusing walls}
\label{3_walkers}

\begin{figure}[htb]
  \begin{center}
\includegraphics[width=0.5\textwidth]{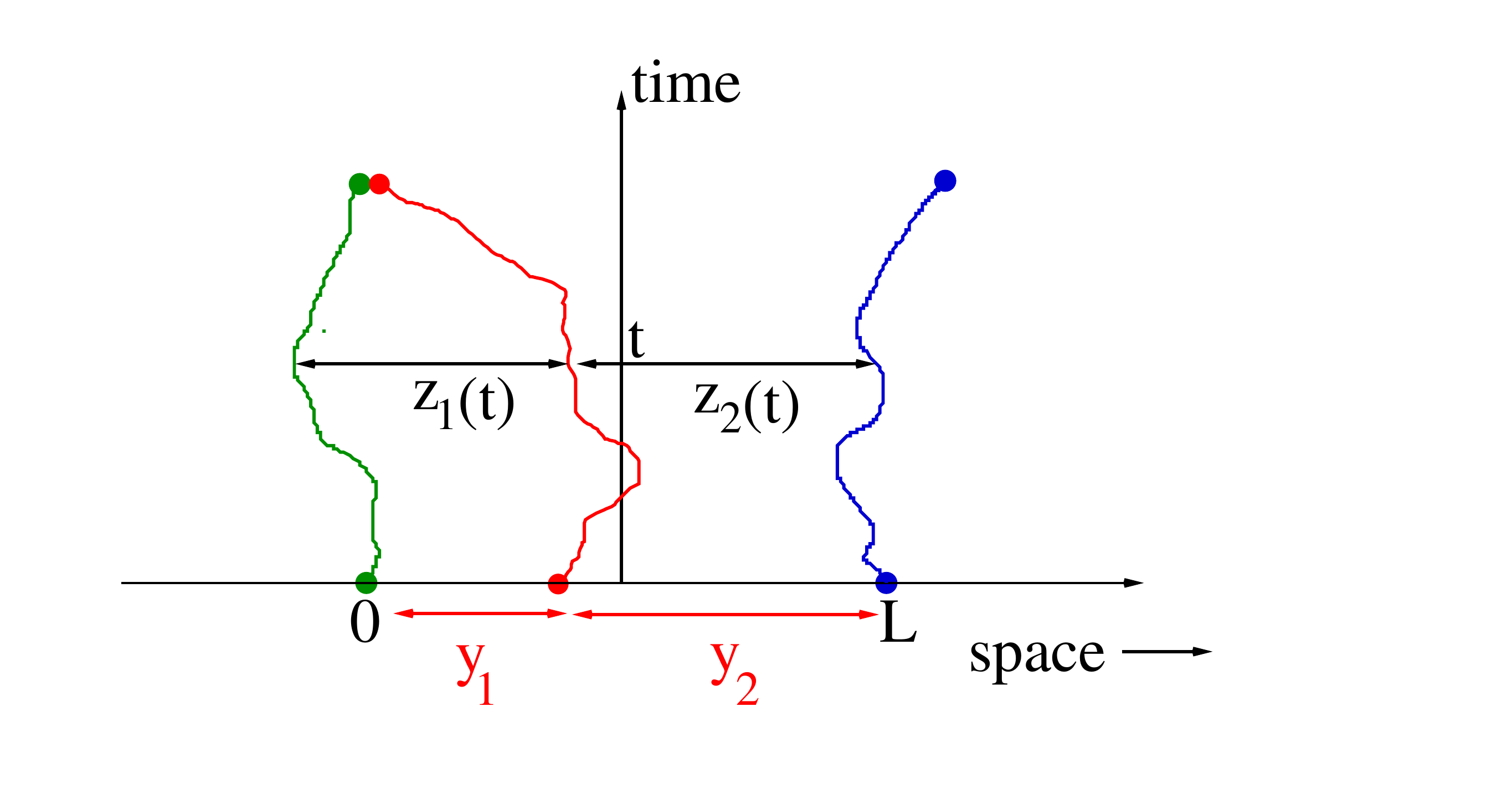}
  \end{center}
\caption{A schematic trajectory of a single Brownian particle $x_2(t)$ (shown by the red solid line)
diffusing between two moving walls: the left wall $x_1(t)$ starts at 0 (its trajectory shown by the green solid line)
and the right wall $x_3(t)$ starts at $L$ (with a blue trajectory). The initial gaps are denoted
by $0\le y_1\le L$ and $0\le y_2\le L$, with $y_1+y_2=L$. The process
stops when the middle walker  hits either the left boundary  or
the right boundary. In the displayed configuration,
the walker hits the left boundary before hitting the right boundary.
We also denote the gaps between the positions
of the walkers by $z_1(t)=x_2(t)-x_1(t)$ and $z_2(t)=x_3(t)-x_2(t)$ at time $t$ before the process stops.}
\label{fig:3bm}
\end{figure}

We next consider a generalized model in one dimension where a single Brownian walker diffuses
in the region sandwiched between between two walls as above, except that the walls themselves
diffuse in time. In this section we introduce the model formally.
Thus, we now consider three Brownian walkers on a line whose 
positions $x_i(t)$ ($i=1,2,3$)
evolve via independent Langevin equations
\begin{equation}
\frac{dx_i}{dt}= \sqrt{2D_i}\, \eta_i(t)\, ,
\label{3bm_lange.1}
\end{equation}
where $D_i$'s are the respective diffusion constants and $\eta_i(t)$'s are zero mean Gaussian white noises
with correlator 
\begin{equation}
\langle \eta_i(t)\eta_j(t')\rangle= \delta_{i,j}\, \delta(t-t')\, .
\label{noise_corr.1}
\end{equation}
The walkers start from ordered initial positions satisfying $x_1(0)<x_2(0)<x_3(0)$. The process stops
when the middle walker hits either of the moving boundaries on the left or the right 
(see Fig.~\ref{fig:3bm}). 
Let $y_1= x_2(0)-x_1(0)$ and $y_2= x_3(0)-x_2(0)$ 
denote the initial gaps, as in Section II for the case of fixed walls.
Let us denote the
the splitting probabilities by $p_{\rm left}(y_1,y_2)$ and $p_{\rm right}(y_1,y_2)$ for first hitting the left and right walls, respectively. This
is equivalent to the definition Eq.~\eqref{def:p:left:right},
but now depends explicitly on two distances,
which we will use below for the analytical calculation.
Evidently, $p_{\rm left}(y_1,y_2)+ p_{\rm right}(y_1,y_2)=1$. Note that when $D_1=D_3=0$, this problem 
reduces precisely to the fixed-wall problem with $D=D_2$
as studied in Section \ref{one:walker}. Below, we first describe the numerical 
approach in Section \ref{Num_section} and then present the exact analytical
result in Section \ref{Ana_section}.

\subsection{Numerical approach to compute the splitting probability}
\label{Num_section}

Since we compare our analytical results with numerical simulations~\cite{practical_guide2015} throughout the paper, 
we first briefly describe the 
simulation procedure.
The basic idea is to simulate Brownian motions for the walkers
$x_i(t)$, $i=1,2,3$. 
For simplicity, let us first consider Eq.~(\ref{3bm_lange.1}) with the initial condition 
$x_i(0)=0$. Then the position distributions at time $t$ are given by
\begin{equation}
p_t(x_i)=\frac 1 {\sqrt{4 \pi D_i t}} \exp\left(-\frac{x_i^2}{4  D_i t}\right)\,
\end{equation}
i.e., a Gaussian with zero mean and standard deviation $\sqrt{2 D_i t}$.
If the walkers start at non-zero initial positions $x_i(0)$, the
probability distributions are correspondingly shifted.

This solution is true also for small time intervals of size $\Delta t$,
which can be interpreted as the relative movement, i.e., a step,
of a walker within
the small time interval, as utilized within a numerical simulation.
Summing up several steps leads to the walk.
This is described by a
discretised version of the Langevin equation for a time series
$x(0)$, $x(\Delta t)$,
$x(2\, \Delta t), \ldots$ at discretised times $t=k \Delta t$
($k=0,1,2,\ldots k_{\max}$) until the final time $t_{\max}=k_{\max} \Delta t$.
and reads 
\begin{equation}
x_i((k+1)\Delta t) = x_i(k \Delta t) + \sqrt{2 D_i \Delta t} \xi_i(k)\,,
\label{eq:step:RW}
\end{equation}
with indendependent and identically distributed (iid) random numbers $\xi_i(k)$ which are Gaussian distributed
with zero mean and variance one.
Note that these equations do not carry any discretisation
error at the measured times $t=k\Delta t$.

Since one main interest of this study is first-passage or first-splitting
properties, this
will lead to approximations: between the measured positions
$x_i(k \Delta t)$ the behavior 
is not known precisely, except that the paths will not be ``far away''
from the measured
positions, respectively.
For the average behavior, if $\Delta t$ is chosen small enough,
this will not matter much, as we see for our results.
We have used mostly $\Delta t=10^{-5}$ in this work and did tests
for $\Delta t=10^{-4}$ which did not significantly change the results.
Thus, for those simulations, where the final time and the
number of independent runs are large, we used $\Delta t=10^{-4}$.
Note that a more efficient and accurate but more complex
algorithm with adaptive 
time steps can certainly  be devised. This has already been
done for the first-passage time properties of a single Brownian
walker with a single non-moving target
\cite{walter2020}. The main idea of that algorithm is to
adaptively splitting those time
intervals where with large-enough probability a hitting event occurs.
Such an algorithm would, in particular, be useful if one is interested in
short time properties. Here,
as mentioned, we simply choose $\Delta t$ small enough to
measure our quantities of interest. As we will see below, we observe very
good agreement between numerical and analytical results, which shows
that the time steps we have chosen are small enough.

One main quantity of interest is the left-hitting probability
$p_{\rm left}(y_2|L)$
which depends on the initial distance $y_2=x_3(0)-x_2(0)=L-x_2(0)$
of the middle to the right
walker, i.e., on $x_2(0)$. One
could perform simulations for many values of $y_2 \in [0,L]$ but that
would be tedious. Instead, we generate just three random walks and
check a set  $0<l_1<l_2<\ldots < l_N<L$ of $N$
starting positions for the middle walker $x_2(0)$. Here we consider
equally spaced position $l_n=n\Delta x$ for suitable small spatial
resolution $\Delta x$, here $\Delta x=L/(N+1)$, typically we use $N=50$. 
Now, one \emph{run} consist of generating
three time series $\{x_i(t)\}$ with $x_1(0)=x_2(0)=0$ and $x_3(0)=L$.
The output of the analysis consist of two vectors $\tilde p_{\rm left}(l_n)\in \{0,1\}$
and $\tilde p_{\rm right}(l_n)\in \{0,1\}$ ($n=1,\ldots,N$)
which tell whether for given set of three time series,
the middle walker
starting at the position $y_2=l_n$,  leads to hitting the left or right
moving wall first, respectively.
These checks can be done rather quickly, because for each time $t$
one has to check only an interval $I=[l_{\rm left}, l_{\rm right}]$ of
starting positions $x_2 \in I$, where $I$ is always chosen such that
neither hitting the left nor the right boundary
has been observed until the current time $t$. Hence, initially,
ate time $t=0$, one has $I=[l_1,l_N]$.
With increasing time $t$, more and more hitting events for some
starting positions of $x_2$ 
are observed and the interval $[l_{\rm left}, l_{\rm right}]$ shrinks. The
check is finished when $I$ is empty, i.e., for all considered
values of $x_2(0)$ a hit of the middle walker with  a boundary has
been observed. This process achieved is by the following algorithm:

\mbox{
\begin{minipage}{0.9\textwidth}
\begin{tabbing} xx \= xx \= xx \= xx \= xx \= xx \= xx \= xx \= xxxxxxxxxx \=
   xx \= x \= xx \kill
 {\bf algorithm} {\bf check\_boundary($t_{\max}$, $\{x_i(t)\}$, $\{l_i\}$)}\\
 {\bf begin}\\
 \> initialize all $\tilde p_{\rm left}(l_n)=\tilde p_{\rm right}(l_n)=0$;\\
 \> set $t=0$; $l_{\rm left}=l_1$, $l_{\rm right}=l_N$\\
 \> {\bf while} ($t<t_{\max}$ AND $l_{\rm left}< l_{\rm right}$)\\
 \>\> {\bf for} $l=l_{\rm left} \ldots l_{\rm right}$
 \>\>\>\>\>\>\> (iterate start) \\
 \>\> {\bf begin}\\
 \>\>\>  {\bf if} ($l+x_2(t) < x_1(t)$) {\bf then}
 \>\>\>\>\>\>\>\> (hit left) \\
 \>\>\>\> set $\tilde p_{\rm left}(l)=1$; $l_{\rm left}=l+\Delta x$; \\
 \>\>\>  {\bf if} ($l+x_2(t) > x_3(t)$) {\bf then}
 \>\>\>\>\>\>\> (hit right)\\
 \>\>\> {\bf begin}\\
 \>\>\>\> set all $\tilde p_{\rm left}(l)=1$ $\ldots$
 $\tilde p_{\rm left}(l_{\rm right})=1$; \\
 \>\>\>\> $l_{\rm right}=l-\Delta x$; \\
 \>\>\> {\bf end}\\
 \>\>\> $t = t +\Delta t$;\\
 \>\> {\bf end}\\
 \> {\bf return} ($\{ \tilde p_{\rm left}(l_n), \tilde p_{\rm right}(l_n)\}$);\\
 {\bf end}\\
\end{tabbing}
\end{minipage}
}

When averaging over a certain number of runs, we usually did
$10^4$, the averages of
$\tilde p_{\rm left}(l_n)$ and $\tilde p_{\rm left}(l_n)$ will result
for the considered values $y_2=L-l_n$ ($n=1,\ldots,N$)
in estimates for $p_{\rm left}(y_2|L)$  and
$p_{\rm right}(y_2|L)$. Note that due to the discretisation in time,
it might happen that for a run and a given starting position $l_n$
one observes hits both with  the left and right
boundary at the same time, within the discretisation
accuracy. For a test with $\Delta t=10^{-5}$
it happened never in $10^4$ runs, but
for $\Delta t =10^{-4}$, which we used sometimes,
it happened just  16 times, combined for all possible starting positions
together. This has no visible effect on the results.

As mentioned earlier,  we considered the movements only until a maximum time $t_{\max}$.
This leads to very few cases for some values of the initial positions $x_2(0)=l_n$
for which a small fraction of trajectories remain undecided, as visible from
$\tilde p_{\rm left}(l_n)=\tilde p_{\rm left}(l_n)=0$ after the algorithm has
finished. For our choice $L=1$ and
$t_{\max}=10$ this happens for a fraction of at most $10^{-3}$
of the runs, which is not visible in the $p_{\rm left}(y_2|L)$ plots.

We also measure numerically
the joint distribution of times $t$ of the middle walker hitting the left
boundary and the corresponding distances
$z=x_3(t)-x_2(t)$ from the right boundary.
Here, we have restricted ourselves to the middle starting
position $x_2(0)=L/2$ of the middle walker to save the amount of stored
data.

The data generated in
this work are publicly
available in the DARE repository of the University of Oldenburg
\cite{data_three_BM}.

\subsection{Exact result for the splitting probability}
\label{Ana_section}

We first introduce the gap variables $z_1(t)=x_2(t)-x_1(t)$ and $z_2(t)=x_3(t)-x_2(t)$.
Using Eq.~(\ref{3bm_lange.1}) it is easy to see that they evolve via the pair of Langevin equations
\begin{eqnarray}
\frac{dz_1}{dt}&= &\sqrt{2D_1}\, \eta_1(t)-\sqrt{2D_2}\, \eta_2(t)= \xi_1(t) \label{z1_lange} \\
\frac{dz_2}{dt} & =& \sqrt{2D_3}\, \eta_3(t)-\sqrt{2D_2}\, \eta_2(t)= \xi_2(t) \label{z2_lange}
\end{eqnarray}
where $\xi_{1,2}(t)$ are again zero-mean Gaussian white noises, but they are correlated
since they share the same noise $\eta_2(t)$. The correlators of these noises are given by
\begin{eqnarray}
\langle \xi_1(t)\xi_1(t')\rangle & = & 2(D_1+D_2)\,\delta(t-t') \nonumber \\ 
\langle \xi_2(t)\xi_2(t')\rangle & = & 2(D_2+D_3)\,\delta(t-t') \\
\langle \xi_1(t)\xi_2(t')\rangle & = & - 2 D_2 \delta(t-t')\, . \nonumber
\label{xi_corr.1}
\end{eqnarray}
Thus the noises $\xi_1(t)$ and $\xi_2(t)$ are anti-correlated.
This anticorrelation is expected because
if the central
particle moves to the left (or right), the gap $z_1(t)$ decreases (increases) while the
gap $z_2(t)$ increases (decreases), leading to the anti-correlation. The `two-dimensional'
gap process $(z_1(t), z_2(t))$
starts from the initial condition $(z_1(0)=y_1, z_2(0)=y_2)$. Our goal is to compute 
the splitting probability
$p_{\rm left}(y_1,y_2)$, i.e., the probability of the event when $z_1(t)$ hits zero before $z_2(t)$ does so.

To compute this splitting probability, we again use the backward Fokker-Planck approach using
the Langevin equations (\ref{z1_lange}) and (\ref{z2_lange}). As in the previous subsection,
we now evolve the two-dimensional process $(z_1(t), z_2(t))$ from its initial value $(y_1,y_2)$
in the first step of duration $\Delta t$. In this first step, the
two-dimensional process moves to
$(y_1+ \xi_1(0)\, \Delta t, y_2+ \xi_2(0)\, \Delta t)$ and the process diffuses subsequently starting from
this new position. Here $(\xi_1(0), \xi_2(0))$ refers to the first kick or noise to the process.
Hence it follows that
\begin{equation}
p_{\rm left}(y_1,y_2)= \langle p_{\rm left}(y_1+ \xi_1(0)\, \Delta t, y_2+ \xi_2(0)\, \Delta t)\rangle\, ,
\label{bfp.1}
\end{equation}
where $\langle \ldots \rangle$ refers to the average over the initial noises $(\xi_1(0), \xi_2(0))$.
Expanding
in Taylor series for small $\Delta t$ up to the second order and using the
properties of the noise correlators in Eq.~(\ref{xi_corr.1}), one gets
a two-dimensional partial differential equation in the $(y_1, y_2)$ plane, or
more precisely in the $(y_1\ge 0, y_2\ge 0)$ quadrant
\begin{equation}
(D_1+D_2)\, \frac{\partial^2 p_{\rm left}}{\partial {y_1}^2}
+ (D_2+D_3)\, \frac{\partial^2 p_{\rm left}}{\partial {y_2}^2}
 -2 D_2\, \frac{\partial^2 p_{\rm left}}{\partial y_1\partial y_2}=0\, .
\label{2d_ode.1}
\end{equation}
It satisfies the two boundary conditions: (i) $p_{\rm left}(y_1=0, y_2)=1$ and (ii) $p_{\rm left}(y_1,y_2=0)=0$.
In case (i), if the process starts from $y_1=0$, then the probability that it hits the left boundary before
the right boundary is surely $1$. In contrast, in case (ii), it the initial value $y_2=0$, the central
walker surely hits the right boundary before the left and hence $p_{\rm left}(y_1,y_2=0)=0$.

\begin{figure}[htb]
\includegraphics[width=0.48\textwidth]{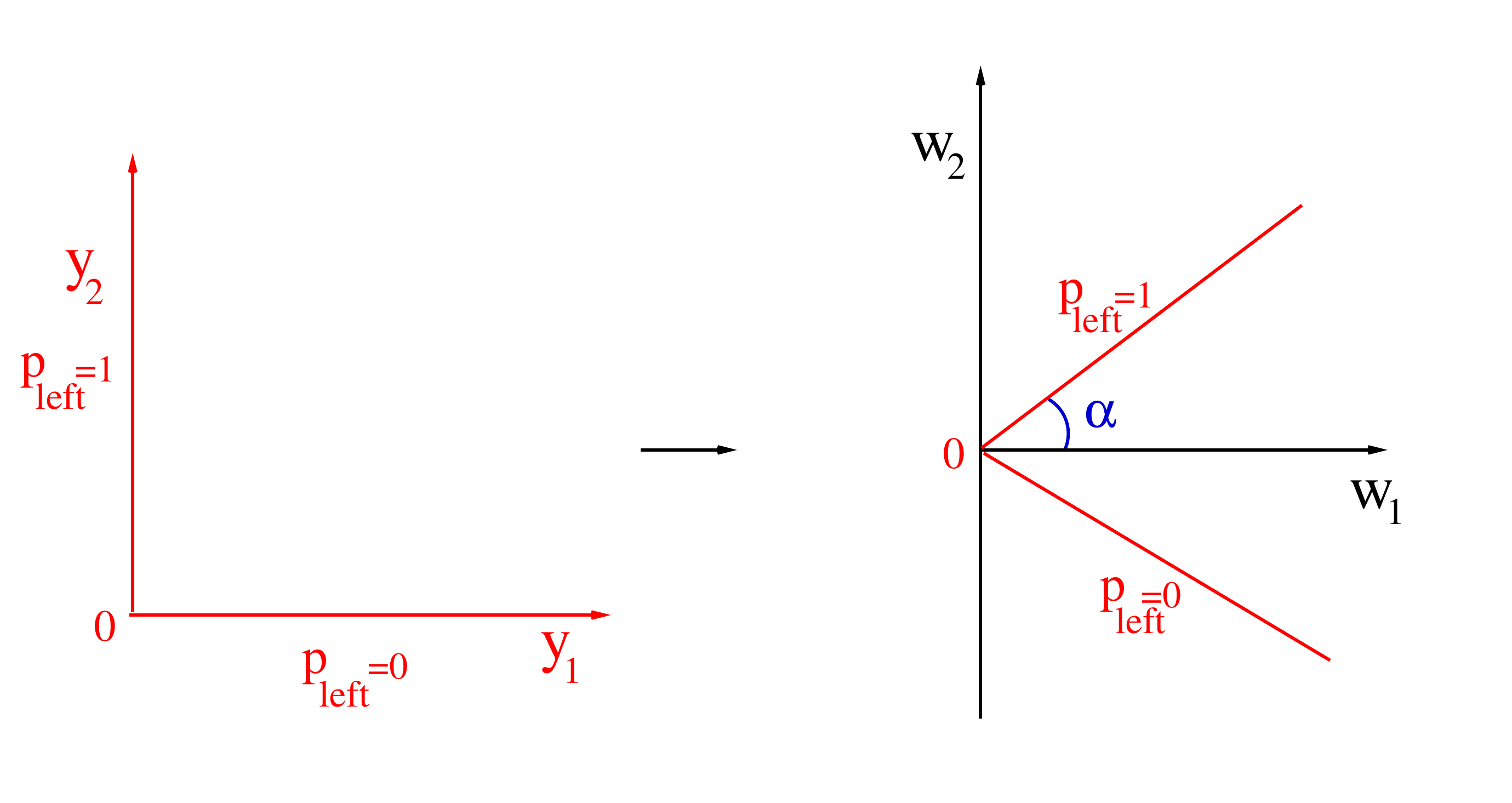}
\caption{Under the change of variables in Eq.~(\ref{w.1}), the quadrant $(y_1\ge 0, y_2\ge 0)$ in
the $(y_1,y_2)$ plane maps onto a wedge in the $(w_1,w_2)$ plane, bounded by the two lines
$w_2= \pm \tan(\alpha)\, w_1$ with wedge angle $\alpha_W=2\alpha$. The boundary conditions 
$p_{\rm left}(y_1=0,y_2)=1$
and $p_{\rm left}(y_1, y_2=0)=0$ in the $(y_1, y_2)$ plane accordingly translate
into the boundary conditions on the two lines bounding the wedge in the $(w_1,w_2)$ plane.}
\label{fig:wedge}
\end{figure}

To solve Eq.~(\ref{2d_ode.1}) it is convenient first to make a change of variables $(y_1, y_2)\to (w_1, w_2)$
to get rid of the off-diagonal term $-2\,D_2\, \frac{\partial^2 p_{\rm left}}{\partial y_1\partial y_2}$
in Eq.~(\ref{2d_ode.1}) and reduce it to a standard isotropic Laplace's equation. 
The same change of variables was used before in the context of computing the distribution
of the maximum displacement between the leader and the laggard in the $3$-walkers problem~\cite{MB2010}.
This amounts to
essentially diagonalize the $(2\times 2)$ matrix with elements $[{D_1+D_2, -D_2},{-D_2, D_2+D_3}]$.
The change of variable that achieves this diagonalization can be written as
\begin{eqnarray}
w_1 & =& \frac{\sqrt{D_0}}{\sqrt{2-\gamma}}\, \left[ \frac{y_1}{\sqrt{D_1+D_2}}+
\frac{y_2}{\sqrt{D_2+D_3}}\right] \nonumber \\
w_2 &=& \frac{\sqrt{D_0}}{\sqrt{2+\gamma}}\, \left[ -\frac{y_1}{\sqrt{D_1+D_2}}+
\frac{y_2}{\sqrt{D_2+D_3}}\right] 
\label{w.1}
\end{eqnarray}
where we have defined
\begin{equation}
\gamma= \frac{2D_2}{\sqrt{(D_1+D_2)(D_2+D_3)}}\, .
\label{gamma_def}
\end{equation}
In Eq.~(\ref{w.1}), we have included a constant $D_0$ in order that $w_i$'s have the same dimensions
(length) as the $y_i$'s. One can choose any positive value of $D_0$ and we will see shortly
that the splitting probability $p_{\rm left}(y_1,y_2)$ is independent of the actual value of 
$D_0$, as long as it is nonzero and positive. Under this change of variables, Eq.~(\ref{2d_ode.1})
simply reduces to the Laplace's equation in the $(w_1, w_2)$ plane
\begin{equation}
D_0\, \left[ \frac{\partial^2 p_{\rm left}}{\partial w_1^2}+ 
\frac{\partial^2 p_{\rm left}}{\partial w_2^2}\right]=0\, ,
\label{laplace.1}
\end{equation}
which shows immediately that $D_0$ drops out. The lines
$y_1=0$ and $y_2=0$ in the $(y_1,y_2)$ plane, under the change of variables \eqref{w.1}, translate
into a pair of straight lines $w_2=\pm \tan(\alpha)\, w_1$  
where
\begin{equation}
\alpha= \tan^{-1}\left(\sqrt{\frac{2-\gamma}{2+\gamma}}\right)\, .
\label{angle.1}
\end{equation}
Hence, the quadrant $(y_1\ge 0, y_2\ge 0)$ transforms into the wedge in the $(w_1,w_2)$ plane
bounded by the straight lines $w_2=\pm \tan(\alpha)\, w_1$ (see Fig.~\ref{fig:wedge}).
The wedge angle is $\alpha_W=2\alpha$.
Consequently, we need to solve the Laplace's equation (\ref{laplace.1}) in this wedge
in the $(w_1,w_2)$ plane with boundary conditions: (i) $p_{\rm left}(w_1,w_2)=1$ along
$w_2= \tan(\alpha)\, w_1$ and (ii) $p_{\rm left}(w_1,w_2)=0$ along
$w_2= -\tan(\alpha)\, w_1$.

The wedge geometry naturally suggests that it would be easier to solve the Laplace's equation (\ref{laplace.1})
in polar coordinates $(w_1,w_2)\to (r,\phi)$, where it reads
\begin{equation}
\frac{\partial^2 p_{\rm left}}{\partial_r^2} +\frac{1}{r}\, \frac{\partial p_{\rm left}}{\partial r}
+\frac{1}{r^2}\, \frac{\partial^2 p_{\rm left}}{\partial \phi^2}=0\, ,
\label{polar.1}
\end{equation}
valid for $r\ge 0$ and $-\alpha\le \phi \le \alpha$ with boundary conditions: (i) $p_{\rm left}(r, \phi=\alpha)=1$
and $p_{\rm left}(r, \phi=-\alpha)=0$ for all $r\ge 0$. The fact that the boundary conditions must hold
for any $r\ge 0$ indicates that the solution must be independent of $r$. In fact, assuming $p_{\rm left}(\phi)$
to be independent of $r$ shows that it must satisfy
\begin{equation}
\frac{d^2 p_{\rm left}(\phi)}{d \phi^2}=0\, ,
\label{ode_polar.1}
\end{equation}
whose general solution is $p_{\rm left}(\phi)= a_1 + b_1\, \phi$. The boundary conditions (i) and (ii) fixes
the two unknown constants $a_1$ and $b_1$ and we obtain the full solution
\begin{equation}
p_{\rm left}(\phi)= \frac{1}{2}\left(1+ \frac{\phi}{\alpha}\right)\, ,
\label{sol.1}
\end{equation}
where the angle $\alpha$ is given in Eq.~(\ref{angle.1}). We note that the polar angle is given by 
\begin{equation}
\phi= \tan^{-1}\left(\frac{w_2}{w_1}\right)\, .
\label{sol.2}
\end{equation}
Substituting \eqref{sol.2} in Eq.~(\ref{sol.1}) and transforming back to the original coordinates $(y_1,y_2)$
via Eq.~(\ref{w.1}) we get our desired exact solution for the splitting probability
\begin{eqnarray}
\label{final_sol.0}
p_{\rm left}(y_1,y_2) &=&  \frac{1}{2}
+ \frac{1}{2\, \tan^{-1}\left(\sqrt{\frac{2-\gamma}{2+\gamma}}\right)}
\,  \\
& & \tan^{-1}\left[ \sqrt{\frac{2-\gamma}{2+\gamma}}\, \frac{
\left( -\frac{y_1}{\sqrt{D_1+D_2}}
+ \frac{y_2}{\sqrt{D_2+D_3}}\right)}{\left( \frac{y_1}{\sqrt{D_1+D_2}}
+ \frac{y_2}{\sqrt{D_2+D_3}}\right)}\right]\, , \nonumber
\end{eqnarray}
where $\gamma$ is defined in Eq.~(\ref{gamma_def}). Note that using $y_1+y_2=L$, we can eliminate
$y_1$ in terms of $y_2$ in \eqref{final_sol.0} and express $p_{\rm left}(y_2|L)$ only
in terms of $y_2$ and $L$ in a more compact form. It reads,
\begin{equation}
p_{\rm left}(y_2|L)=  \frac{1}{2}\left[1+ \frac{1}{\tan^{-1}(a)}\, \tan^{-1}\left(
a\, \frac{y_2(b+1)-L}{y_2(b-1)+L}\right)\right]\, ,
\label{final_sol.1}
\end{equation}
for $0\le y_2\le L$,
where the two constants $a$ and $b$ are given by
\begin{equation}
a=\sqrt{\frac{2-\gamma}{2+\gamma}}\quad
{\rm and}\quad b= \sqrt{\frac{D_1+D_2}{D_2+D_3}}\, .
\label{ab_def}
\end{equation}
Near the two limits $y_2\to 0$ and $y_2\to L$, the splitting probability has the asymptotic behaviors
\begin{eqnarray}
p_{\rm left}(y_2|L) \approx \begin{cases}
\frac{a\, b}{(1+a^2)\, \tan^{-1}(a)}\, \frac{y_2}{L} \quad\quad\quad\quad\quad {\rm as}\quad y_2\to 0 \nonumber \\
\\
1- \frac{a}{(1+a^2) b \tan^{-1}(a)}\, \frac{(L-y_2)}{L} \quad\,\, {\rm as} \quad y_2\to L\, .
\end{cases}
\label{asymp.1}
\end{eqnarray}
Thus, unlike the fixed wall case where the splitting probability $p_{\rm left}(y_2|L)$ was
a simple linear function in $y_2\in [0,L]$, the exact result \eqref{final_sol.1} for the
moving walls case is nonlinear and has a richer structure.

\begin{figure}
\includegraphics[width=0.47\textwidth]{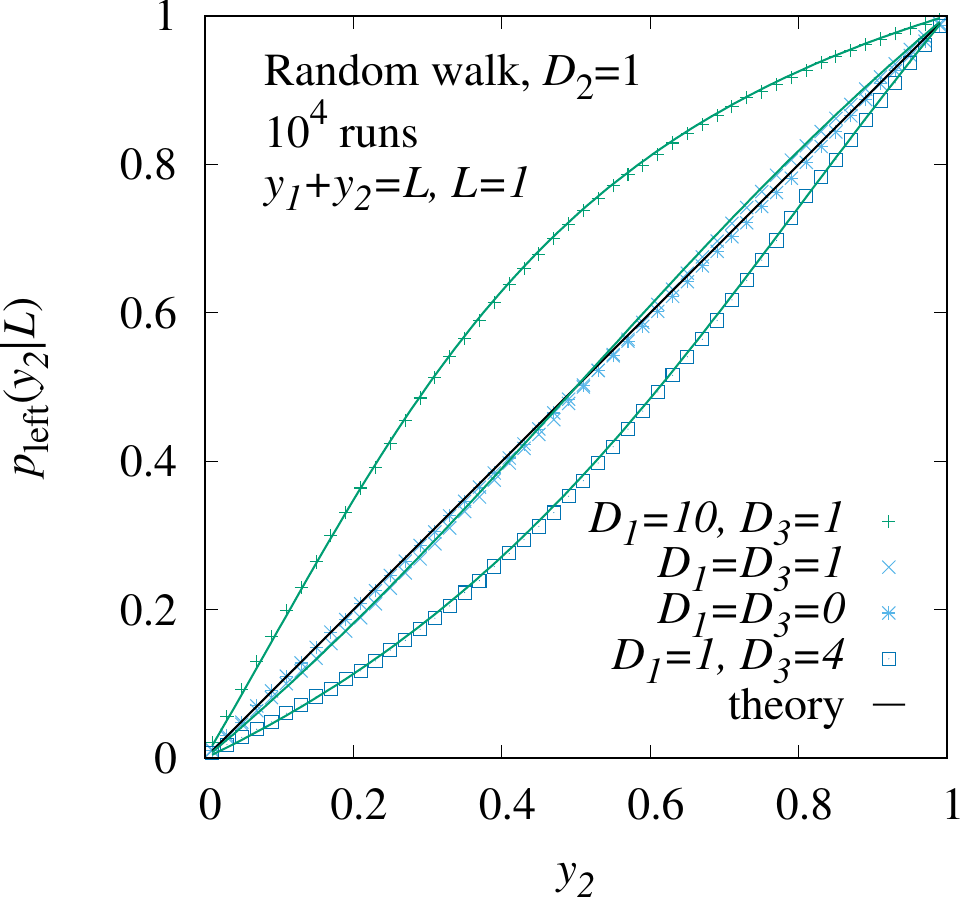}
\caption{Theoretical prediction for the splitting probability $p_{\rm left}(y_2|L)$ as
  a function of $y_2$ in Eq.~(\ref{final_sol.1}), shown as lines in different
  colors,
  compared with numerical simulations, shown as symbols,
  for different choices of $D_1$ and $D_3$ (with $D_2=1$ and $L=1$ fixed).
  One sees excellent agreement between the theory and simulations. }
\label{fig:hp_num}
\end{figure}

As a useful consistency check of the exact result in \eqref{final_sol.1}, we note that for 
$D_1=D_3=0$ (corresponding to fixed walls), we have $\gamma=2$ from
Eq.~(\ref{gamma_def}). Taking the $\gamma\to 2$ limit
in Eq.~(\ref{final_sol.1}) gives a linear curve
\begin{equation}
p_{\rm left}(y_2|L)= \frac{y_2}{L}\, , \quad   0\le y_2\le L\, .
\label{limit_sol.1}
\end{equation}
We thus recover, as expected, the solution \eqref{splitting_BM.1} for the fixed wall case. Another interesting
special case corresponds to $D_1=D_2=D_3=D$. In this case, we get $\gamma=1$ from (\ref{gamma_def}).
Using $\tan^{-1}(1/\sqrt{3})= \pi/6$, Eq.~(\ref{final_sol.1}) gives the splitting probability
\begin{equation}
p_{\rm left}(y_1, y_2)= \frac{1}{2}+ \frac{3}{\pi}\, 
\tan^{-1}\left(\frac{2\,y_2-L}{\sqrt{3}\, L}\right)\, .
\label{lim_sol.2}
\end{equation}
In Fig.~\ref{fig:hp_num}, we compare our analytical prediction in
Eq.~(\ref{final_sol.1}) 
with numerical simulations for several choices of $(D_1, D_2, D_3)$,
using $t_{\max}=10$, averaged over $10^4$ independent runs,
finding excellent agreement. For all results we have used the same
diffusion $D_2=1$ for the middle walker. For fixed boundaries, i.e.,
$D_1=D_3=0$, the behavior follows the straight line Eq.~(\ref{splitting_BM.1}).
The stronger the diffusion of the walks, i.e., the larger $D_1$ and $D_2$,
the stronger are the deviations from the straight line.

\begin{figure}
\includegraphics[width=0.47\textwidth]{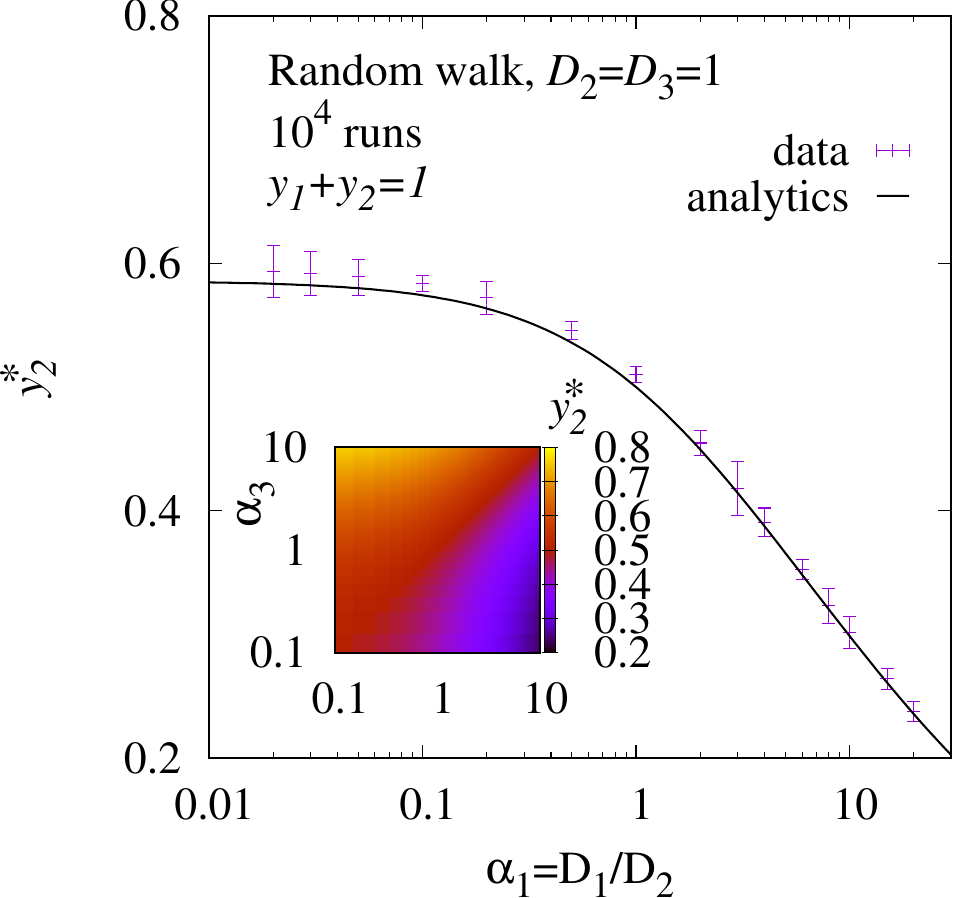}
\caption{Theoretical prediction for the halfpoint $y_2^*(\alpha_1)$ (where the left and right splitting probabilities
are exactly equal) as a function of $\alpha_1$ in \eqref{symm.2} is shown by the solid black line.
The numerical data points are shown by $+$ symbols. In the inset, we show a heat map
of $y_2^*$ in \eqref{symm.1} in the $(\alpha_1,\alpha_3)$ plane (we have set $L=1$).} 
\label{fig:halfpoint}
\end{figure}

Finally, we make an interesting observation. In the case of fixed walls $(D_1=0, D_2, D_3=0)$,
it is clear from \eqref{splitting_BM.1} that exactly at the midpoint $y_2=y_2^*=L/2$, the splitting probabilities
$p_{\rm left}=p_{\rm right}=1/2$. This just follows from symmetry. However, for moving walls with
nonzero diffusion constants $D_1$ and $D_3$,
there is no obvious symmetry, and consequently the splitting probabilities
are generally different from $1/2$ at the midpoint, as visible in Fig.~\ref{fig:hp_num}.
It is then natural to ask for what value of $y_2=y_2^*$, the left and right
splitting probabilities are exactly equal to each other in the general case? Setting $p_{\rm left}(y_2^*|L)=1/2$
in the exact result \eqref{final_sol.1}, it follows that this symmetrical point occurs at
\begin{eqnarray}
\label{symm.0}
  y_2^* = \frac{L}{b+1}\, , & {\rm and} &  y_1^*=L-y_2^*= \frac{b\, L}{b+1}\, ; \\
& {\rm where} & 
b= \sqrt{\frac{D_1+D_2}{D_2+D_3}}\, . \nonumber
\end{eqnarray}
In terms of the ratios $\alpha_1=D_1/D_2$ and $\alpha_3=D_3/D_2$, the halfpoint
$y_2^*$ in \eqref{symm.0} reads
\begin{eqnarray}
  y_2^* & = & \frac{\sqrt{D_2+D_3}}{\sqrt{D_1+D_2} + \sqrt{D_2+D_3}}\, L
  \nonumber \\
  & = &
\frac{\sqrt{1+\alpha_3}}{\sqrt{1+\alpha_1}+\sqrt{1+\alpha_3}}\, L\, .
\label{symm.1}
\end{eqnarray}
For $D_1=D_3=0$, we clearly recover the fixed wall result $y_2^*=y_1^*=L/2$. 
We verify this analytical prediction via numerical simulation in Fig. \ref{fig:halfpoint},
where, in the main figure, we have set $D_2=D_3=1$ and $L=1$.
In this case $\alpha_3=1$ and $\alpha_1=D_1$. 
From Eq.~(\ref{symm.1}), setting $\alpha_3=1$, the halfpoint $y_2^*(\alpha_1)$ as a function of $\alpha_1$ 
is given by
\begin{equation}
y_2^*(\alpha_1)= \frac{\sqrt{2}}{\sqrt{2}+\sqrt{1+\alpha_1}}\,  .
\label{symm.2}
\end{equation}
In the main curve of Fig. \ref{fig:halfpoint}, we compare this theoretical prediction 
with numerical simulations, finding excellent agreement. The inset in Fig. \ref{fig:halfpoint}
shows a heat map of $y_2^*$ in Eq.~(\ref{symm.1}) in  the $(\alpha_1, \alpha_3)$ plane.

\section{Joint distribution of the hitting time and the separation}
\label{JPDF_tz}

\begin{figure}[htb]
\includegraphics[width=0.48\textwidth]{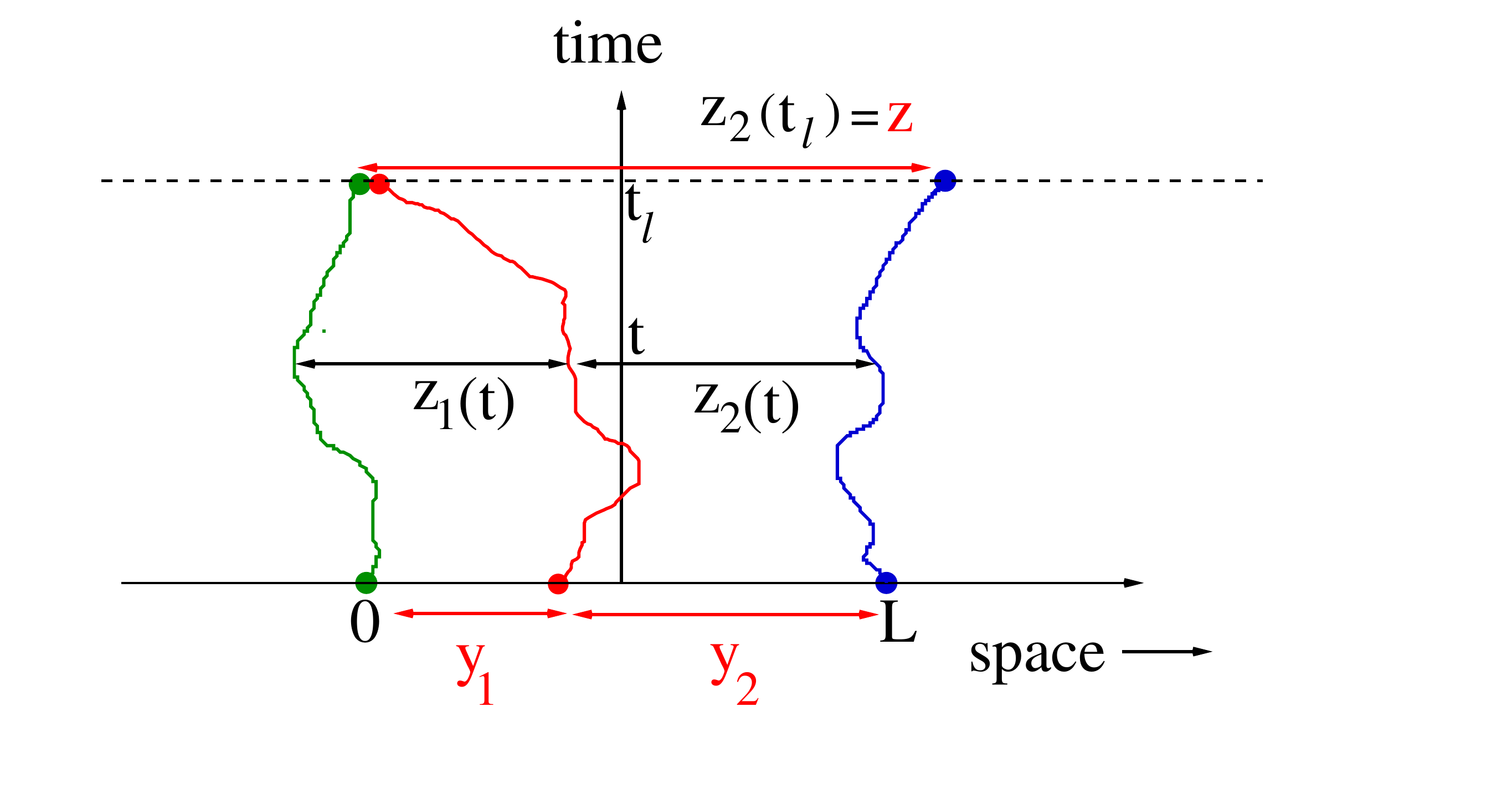}
\caption{A schematic trajectory of a single Brownian particle $x_2(t)$ (shown by the red solid line)
  diffusing between two diffusing walls: the left wall $x_1(t)$ starts at 0
  (green solid line)
and the right wall $x_3(t)$ starts at $L$ (blue solid line). The initial gaps are denoted
by $0\le y_1\le L$ and $0\le y_2\le L$, with $y_1+y_2=L$. We will consider the conditioned
process where the process terminates by the central particle hitting the left boundary first.
In this figure, we show a realization of the conditioned process where $t_{\ell}$ denotes the time of completion and 
$z_2(t_l)=z$ denotes the gap between the central particle and the right boundary at the time of completion.
We also denote the gaps between the positions
of the walkers by $z_1(t)=x_2(t)-x_1(t)$ and $z_2(t)=x_3(t)-x_2(t)$ at time $t$ before the process stops.}
\label{fig:bmfp}
\end{figure}

In the previous section, we studied the splitting probabilities $p_{\rm left}(y_2|L)$
and $p_{\rm right}(y_2|L)= 1-p_{\rm left}(y_2|L)$ as defined
in Eq.~\eqref{def:p:left:right}. 
In this section, we study two other observables in this process. Conditioned on the event that
the process terminates by hitting the left boundary, let
$t_{\ell}$ denote the time at which this hapens (see Fig.~\ref{fig:bmfp}).
Also, at the hitting time of the left boundary $t_{\ell}$, we denote by 
$z_2(t_{\ell})=z$ the separation between the central particle and the right boundary (see Fig.~\ref{fig:bmfp}).
Clearly, both $t_\ell$ and $z_2(t_\ell)=z$ are random variables as they fluctuate from one realization
to another.
Moreover they are generally correlated. The main goal of this section is to compute the
joint distribution of these two random variables, defined as
\begin{align}
  P^{\rm left}(t,z)\, dt\, dz  = \\
  {\rm Prob.}\left[ t_\ell\in [t, t+dt]\,
 {\rm and}    \, z_2(t_\ell)\in [z, z+dz]\right]\, .
\label{joint_tz.1}
\end{align}
This conditional  joint distribution is normalized to unity, i.e.,
$\int_0^{\infty}\int_0^{\infty} P^{\rm left}(t, z)\, dt\, dz=1$.
From this normalized joint distribution, we can then derive the marginal distributions of the random variables 
$t_\ell$ and $z_2(t_\ell)$ by integrating out the other variable. We will denote these marginal distributions
as follows
\begin{eqnarray}
P^{\rm left}_{t_\ell}(t) &= & \int_0^{\infty} P^{\rm left}(t,z)\, dz \,,  \label{margt.1} \\
P^{\rm left}_{z}(z) &=& \int_0^{\infty} P^{\rm left}(t, z)\, dt\, . \label{margz.1}
\end{eqnarray}
Of course, the joint distribution $P^{\rm left}(t,z)$ also depends on the initial gaps $y_1$ and $y_2$, but
we suppress this dependence in $P^{\rm left}(t,z)$ for notational simplicity.

For the computation of the splitting probabilities in the previous section, it was convenient to
use the backward Fokker-Planck approach in Eq.~(\ref{2d_ode.1}) where the initial gaps $y_1$ and $y_2$ were treated
as variables. This was possible because we were not interested there on the actual completion time at which
the central particle hits either the left or the right boundary. However, if we want to compute
the distributions of $t_\ell$ and $z_2(t_\ell)$, we need to keep track of the trajectories as time progresses.
Hence, here we need to employ the forward Fokker-Planck approach where the separations $(z_1(t), z_2(t))$
at time $t$ will be treated as variables. Let $P(z_1,z_2,t)$ denote the probability distribution of $(z_1(t), z_2(t))$
at time $t$ (before the process terminates). Then, from the Langevin equations
\eqref{z1_lange} and \eqref{z2_lange}, one can derive the forward Fokker-Planck equation by increasing
the time from $t$ to $t+\Delta t$ and keeping track of how the distribution evolves. More precisely,
\begin{equation}
P(z_1,z_2,t+\Delta t) = \langle P(z_1- \xi_1(t)\, \Delta t, z_2- \xi_2(t)\, \Delta t, t)\rangle\, ,
\label{ffp.1}
\end{equation}
where $\langle \rangle$ denotes the average over the noises $(\xi_1(t), \xi_2(t))$ at time $t$. 
Note that in order
to arrive at $(z_1,z_2)$ at time $t+\Delta t$, the process must have been at 
$(z_1- \xi_1(t) \Delta t, z_2-\xi_2(t) \, \Delta(t))$
at time $t$ and then reaches $(z_1,z_2)$ by receiving the stochastic increments
$(\xi_1(t), \xi_2(t))$ at time $t$. Finally, since these increments
are stochastic, one must average over them as in Eq.~(\ref{ffp.1}). Expanding in $\Delta t$ for small $\Delta t$ up
to $O((\Delta t)^2)$ and using the properties of noise correlators in Eq.~(\ref{xi_corr.1}), it is easy to show that $P(z_1,z_2,t)$
satisfies the forward Fokker-Planck equation
\begin{equation}
\frac{\partial P}{\partial t}= (D_1+D_2)\, \frac{\partial^2 P}{\partial z_1^2} + (D_2+D_3)\, \frac{\partial^2 P}{\partial z_2^2}
-2\,D_2\, \frac{\partial^2 P}{\partial z_1 \partial z_2}\, ,
\label{ffp.2}
\end{equation}
valid in the region $z_1\ge 0$ and $z_2\ge 0$, with the absorbing boundary conditions: (i) $P(z_1=0, z_2,t)=0$
and (ii) $P(z_1, z_2=0, t)=0$ and the initial condition
\begin{equation}
P(z_1,z_2,0)= \delta(z_1-y_1)\, \delta(z_2-y_2)\, .
\label{init.1}
\end{equation} 

Note that if we integrate $P(z_1,z_2,t)$ over $z_1\ge 0$ and $z_2\ge 0$, it gives the survival probability
\begin{equation}
S(t)= \int_0^{\infty}\int_0^{\infty} P(z_1,z_2,t)\, dz_1\, dz_2\, ,
\label{surv.1}
\end{equation}
that denotes the probability that the process is still alive at time $t$, i.e., the probability that the central particle does
not hit either of the two boundaries up to time $t$. Integrating Eq.~(\ref{ffp.2}) over $z_1$ and $z_2$ and using the boundary conditions
(i) and (ii) above, one finds
\begin{eqnarray}
  F(t)=-\frac{dS}{dt} & = & (D_1+D_2)\, \int_0^{\infty} dz_2\, \frac{\partial P}{\partial z_1}\Big|_{z_1=0} \nonumber \\
& &   + (D_2+D_3)\,
  \int_0^{\infty} dz_1\, \frac{\partial P}{\partial z_2}\Big|_{z_2=0}\\
  & = & F_{\rm left}(t) + F_{\rm right}(t)\, . \nonumber
\label{sp_fp.1}
\end{eqnarray}
This equation admits a nice physical interpretation. The quantity $F(t)=-dS/dt$ on the left hand side (lhs) denotes the
total probability flux leaving the system through the two boundaries at time $t$ and hence it is exactly the first-passage
probability density $F(t)$. Here $F(t)\, dt$ denotes the probability that the process completes exactly between $[t, t+dt]$
via the central particle hitting either the left or the right boundary. This total probability flux 
out of the system can be 
decomposed into two parts: (a) $F_{\rm left}(t)= (D_1+D_2)\, \int_0^{\infty} dz_2\, \frac{\partial P}{\partial z_1}\Big|_{z_1=0}$
denotes the probability current through the left boundary at time $t$ and (b) $F_{\rm right}(t)=(D_2+D_3)\,
\int_0^{\infty} dz_1\, \frac{\partial P}{\partial z_2}\Big|_{z_2=0}$ denotes the probability flux through the right boundary
at time $t$. Note that the splitting probabilities, studied in the previous section
by the backward Fokker-Planck method, can alternatively be obtained by integrating
over $t$, i.e.,
\begin{eqnarray}
  p_{\rm left}(y_2|L) & = & \int_0^{\infty} F_{\rm left}(t')\, dt'\, ,
  \nonumber \\
  {\rm and}
\quad p_{\rm right}(y_2|L) & = & \int_0^{\infty} F_{\rm right}(t')\, dt'\, .
\label{split_fp.1}
\end{eqnarray}

Hence, if we condition on ending the process by the left boundary, the marginal distribution of the completion 
time $t_\ell$ is given by
\begin{equation}
P^{\rm left}_{t_\ell}(t)= \frac{F_{\rm left}(t)}{\int_0^{\infty} F_{\rm left}(t')\, dt'}=\frac{F_{\rm left}(t)}{p_{\rm left}(y_2|L)}\, .
\label{marg_tell.1}
\end{equation}
Upon dividing by the constant conditioning factor $p_{\rm left}(y_2|L)$, 
it then follows that the joint distribution $P^{\rm left}(t_\ell=t, z_2(t)=z)$ can then be read off
the first term in Eq.~(\ref{sp_fp.1}),
namely,
\begin{eqnarray}
\label{jpdf_left.1}
& &  P^{\rm left}(t, z)= \\
& &  \left[\frac{1}{p_{\rm left}(y_2|L)}\right]\, (D_1+D_2)\,\frac{\partial P(z_1, z_2=z,t)}{\partial z_1}\Big|_{z_1=0}\, ,  \nonumber
\end{eqnarray}
so that the marginal distribution of $t_\ell$ is then given by integrating over $z$, i.e.,
\begin{eqnarray}
  \label{ftl.1}
&&  P^{\rm left}_{t_\ell}(t_\ell=t)   =  \
  \int_0^{\infty} dz\, P^{\rm left}(t,z) \\
  && =  \left[\frac{1}{p_{\rm left}(y_2|L)}\right]\, 
(D_1+D_2)\, \int_0^{\infty} dz_2\, 
  \frac{\partial P(z_1,z_2,t)}{\partial z_1}\Big|_{z_1=0}\, ,
  \nonumber
\end{eqnarray}
which precisely coincides with the definition in Eq.~(\ref{marg_tell.1}).
Similarly, by integrating over $t$, one can obtain the other desired marginal
\begin{eqnarray}
 \label{ztl.1}
  & & P^{\rm left}_z(z_2(t_\ell)=z)= \int_0^{\infty} dt\, P^{\rm left}(t,z)
  \\ &&=
\left[\frac{1}{p_{\rm left}(y_2|L)}\right]\, (D_1+D_2)\, \int_0^{\infty} dt\,
\frac{\partial P(z_1,z,t)}{\partial z_1}\Big|_{z_1=0}\, .
\nonumber
\end{eqnarray}

To solve for $P(z_1,z_2,t)$ in Eq.~(\ref{ffp.2}), we again perform the change of variables 
analogous to Eq.~(\ref{w.1}), 
\begin{eqnarray}
w_1 & =& \frac{\sqrt{D_0}}{\sqrt{2-\gamma}}\, \left[ \frac{z_1}{\sqrt{D_1+D_2}}+
\frac{z_2}{\sqrt{D_2+D_3}}\right] \nonumber \\
w_2 &=& \frac{\sqrt{D_0}}{\sqrt{2+\gamma}}\, \left[ -\frac{z_1}{\sqrt{D_1+D_2}}+
\frac{z_2}{\sqrt{D_2+D_3}}\right]\, ,
\label{wz.1}
\end{eqnarray}
where $\gamma$ is given in Eq.~(\ref{gamma_def}) and $D_0$ is kept arbitrary as in Eq.~(\ref{w.1}). The specific 
value of $D_0$
is not important, it is introduced so that $w_1$ and $w_2$ have the dimensions of length. Let $\tilde{P}(w_1,w_2,t)$
denote the probability density in the $(w_1, w_2)$ coordinates which is simply related to $P(z_1,z_2,t)$ via a constant
Jacobian factor
\begin{eqnarray}
  P(z_1,z_2,t)  & = & J\, \tilde{P}(w_1,w_2,t)\quad {\rm where}
  \\
   J & = & \frac{D_0}{\sqrt{D_1\,D_2+D_2\, D_3+ D_3\, D_1}}\, . \nonumber
\label{jacobian.1}
\end{eqnarray}
Then $\tilde{P}(w_1,w_2,t)$ satisfies the two-dimensional diffusion equation
\begin{equation}
\frac{\partial \tilde{P}}{\partial t}= D_0\left [ \frac{\partial^2 \tilde{P}}{\partial w_1^2}+
\frac{\partial^2 \tilde{P}}{\partial w_2^2}\right]\, 
\label{fpw.1}
\end{equation}
within a wedge in the $(w_1,w_2)$ plane that is bounded by the two lines $w_2=\pm \tan(\alpha)\, w_1$ (see Fig.~\ref{fig:wedge}),
with $\alpha$ defined in Eq.~(\ref{angle.1}). The boundary conditions are absorbing on the two bounding lines, i.e.,
$\tilde{P}(w_1,w_2,t)=0$ when $w_2=\pm \tan(\alpha)\, w_1$. Similarly, the initial condition \eqref{init.1} also
transforms accordingly in the $(w_1,w_2)$ coordinates.

The general solution of the diffusion equation in a wedge with absorbing boundary conditions can be easily obtained
in the polar coordinates $(w_1,w_2)\to (r,\phi)$ where
\begin{equation}
r= \sqrt{w_1^2+ w_2^2}\, ; \quad {\rm and}\quad \phi= \tan^{-1}\left(\frac{w_2}{w_1}\right)\, .
\label{polart.1}
\end{equation}
The explicit solution can be obtained by the method of separation of variables. The solution reads
\begin{widetext}
\begin{equation}
\tilde{P}(r,\phi,t)= \frac{1}{\alpha_W\, D_0\, t}\, \sum_{n=1}^{\infty} \sin\left(\frac{n\pi (\alpha+ \phi)}{\alpha_W}\right)\,
\sin\left(\frac{n\pi (\alpha+ \phi_0)}{\alpha_W}\right)\, e^{- (r^2+r_0^2)/{4D_0 t}}\,
I_{n\pi/\alpha_W}\left(\frac{r\, r_0}{2\, D_0\, t}\right)\, ,
\label{gen_sol.1}
\end{equation}
\end{widetext}
where $\alpha_W= 2\alpha$ is the wedge angle and $I_\nu(z)$ is the modified Bessel function of the first kind with
index $\nu$ and argument $z$. Here $(r_0,\phi_0)$ refers to the initial condition from where the diffusing particle
starts. One can then express the solution $\tilde{P}(w_1,w_2,t)$ in cartesian coordinates $(w_1,w_2)$ by
using the transformation in \eqref{polart.1}.
Substituting this general solution $\tilde{P}(w_1,w_2,t)$ in \eqref{jacobian.1} then yields the full exact solution
$P(z_1,z_2,t)$ at all times $t$. Also, using this full solution in Eq.~(\ref{jpdf_left.1}), one can then, in principle,
compute the joint distribution $P^{\rm left}(t,z)$ for all $t$ and all $z$. However, this calculation
is rather cumbersome and it is
more interesting to extract the asymptotic behavior of $P^{\rm left}(t,z)$ for large $t$ and large $z$.
In fact, as we will see below, there is a scaling regime with $t\to \infty$, $z\to \infty$ but with the
ratio $z/\sqrt{t}$ fixed, where the expression of $P^{\rm left}(t,z)$ takes a particularly simple explicit form.
To derive this scaling behavior for large $t$ and large $z$, we need to first extract
the behavior of $P(z_1,z_2,t)$ for large $t$ (and large $z_2$ and $z_1\to 0$) from the general exact solution \eqref{gen_sol.1} 
and then use
it in Eq.~(\ref{jpdf_left.1}) to compute the scaling behavior $P^{\rm left}(t,z)$ .
This is presented in the subsection below.

\begin{figure}
\includegraphics[width=0.5\textwidth]{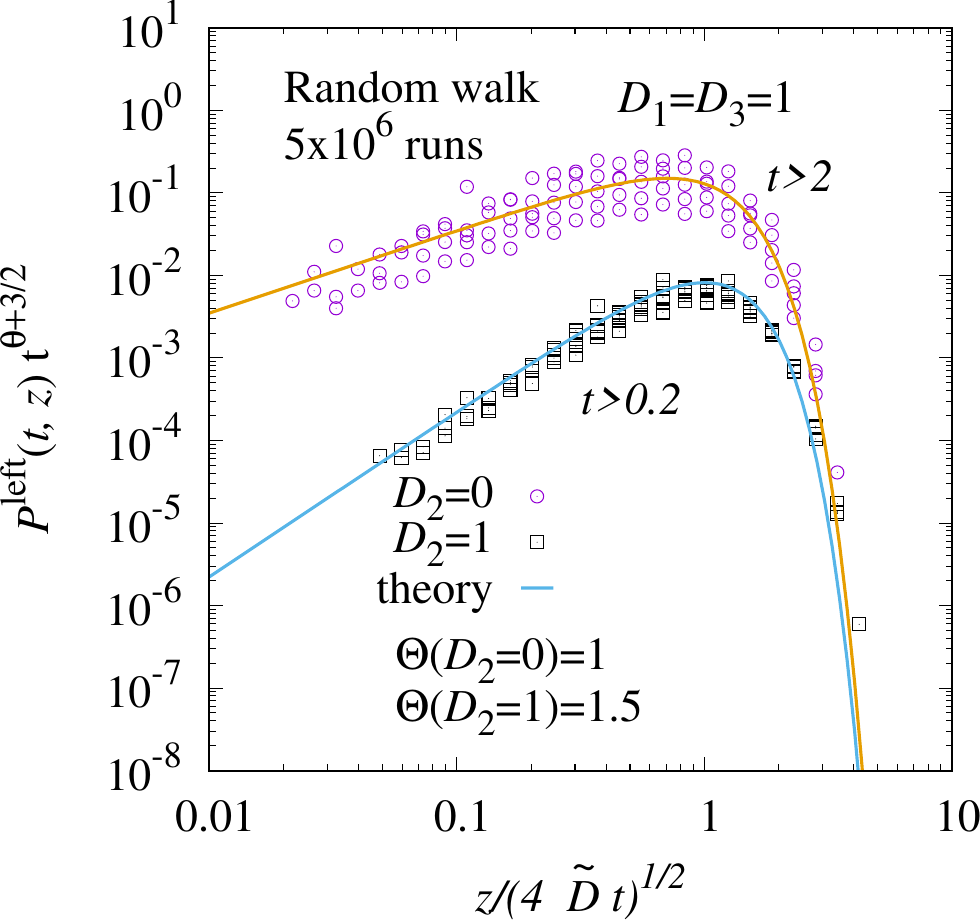}
\caption{Theoretical predictions for the scaling behavior of the
  joint distribution  $P^{\rm left}(t,z)$ in
  Eqs.~(\ref{scaling.1})-(\ref{Hu_def}) shown by solid lines for two
  values of $\Theta$ are compared with  numerical simulations (symbols),
  where the very short-time date is omitted. In
the simulation we fixed $D_1=D_3=1$ and considered two values of $D_2=0$ and $D_2=1$.
In these cases, the theoretical predictions of the exponent $\theta$ in Eq.~
(\ref{theta_def.1}) are respectively $\theta= 1$ (for $D_2=0$) and
$\theta=3/2$ (for $D_2=1$).}
\label{fig:JPDF_tz}
\end{figure}

\subsection{Asymptotic large-$t$ behavior of $P^{\rm left}(z,t)$}

To analyze the large-$t$ behavior of $P(z_1,z_2,t)$, we start from the general solution \eqref{gen_sol.1}
in polar coordinates. We will now consider the scaling limit when $t\to \infty$, $r\to \infty$ with the
ratio $r/\sqrt{t}$ fixed. We keep the starting
radius $r_0\sim O(1)$ fixed.  
We will also use the following asymptotic behavior of $I_\nu(z)$
\begin{equation}
I_\nu(z) \approx \frac{2^{-\nu}}{\Gamma(1+\nu)}\, z^{\nu} \quad {\rm as} \quad z\to 0\, .
\label{Bessel_asymp.1}
\end{equation}
From \eqref{gen_sol.1}, we see that in the scaling regime when $r\sim \sqrt{t}\gg 1$ with $r_0\sim O(1)$, the 
argument $rr_0/(2D_0 t)$ of the Bessel function becomes small. Hence, we can use the asymptotic behavior
in \eqref{Bessel_asymp.1}. This shows that the
dominant contribution to the sum
in Eq.~(\ref{gen_sol.1}) comes from the $n=1$ term, leading to
(after slight rearrangements)
\begin{eqnarray}
  & & \tilde{P}(r,\phi,t)\approx \\
  & & \frac{A\, r_0^{2\theta}}{(D_0\, t)^{\theta+1}}\, \sin\left( \frac{\pi (\alpha+\phi)}{\alpha_W}\right)\,
  e^{-r^2/{4 D_0 t}}\, \left(\frac{r^2}{4\, D_0\, t}\right)^{\theta}\, ,
  \nonumber
\label{polar_late.1}
\end{eqnarray}
where $A$ is an unimportant dimensionless constant dependent on the initial condition (and can be absorbed in the overall normalization
constant) and the exponent $\theta$ is given by
\begin{eqnarray}
\label{theta_def.0}
  \theta & = &
  \frac{\pi}{2\, \alpha_W}\, \quad {\rm where} \\
  \alpha_W & = & 2\, \tan^{-1}\left(\sqrt{\frac{2-\gamma}{2+\gamma}}\,\right)
= \cos^{-1}\left(\frac{\gamma}{2}\right)\, . \nonumber
\end{eqnarray}
Using the expression for $\gamma$ in \eqref{gamma_def}, the exponent $\theta$ reads explicitly
\begin{equation}
\theta= \frac{\pi}{2 \cos^{-1}\left(\frac{D_2}{\sqrt{(D_1+D_2)(D_2+D_3)}}\right)}\, ,
\label{theta_def.1}
\end{equation}
and thus depends continuously on the diffusion constants $\{D_1,D_2,D_3\}$.
In fact, integrating \eqref{polar_late.1} over all space and using \eqref{jacobian.1}, it is clear that the survial probability 
in \eqref{surv.1} decays as a power law at late times
\begin{equation}
S(t) = J\, \int_{-\alpha}^{\alpha} d\phi\, \int_0^{\infty} r\, dr\, \tilde{P}(r,\phi,t) \sim \frac{B_1}{t^\theta}\, ,
\label{surv.2}
\end{equation}
where $B_1$ is a computable constant independent of $t$. The exponent $\theta$ is called the persistence exponent~\cite{BMS2013} and its exact
expression $\theta=\pi/(2\alpha_W)$ in the wedge geometry is well-known in the literature~\cite{Redner_book,BMS2013}.

\begin{figure}
\includegraphics[width=0.47\textwidth]{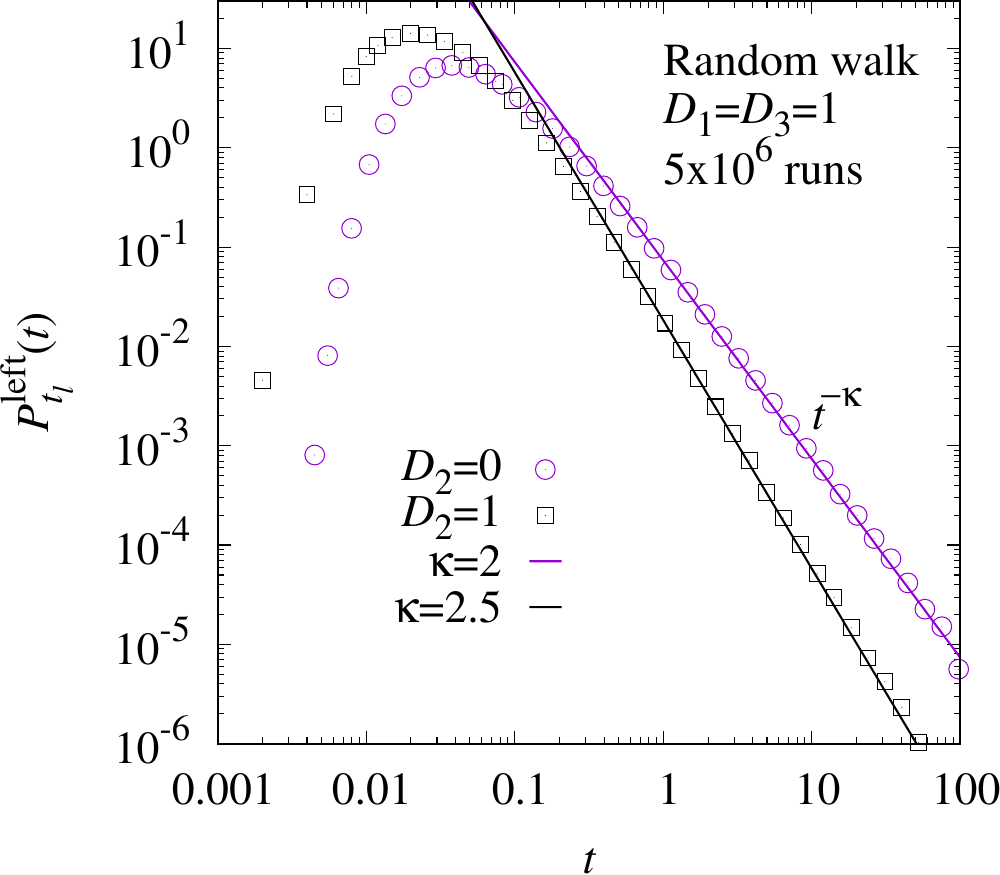}
\caption{Theoretical prediction for the power law decay of the marginal distribution $P^{\rm left}_{t_\ell}(t)
\sim t^{-\kappa}$ in
Eq.~(\ref{margt.asymp}) as
a function of $t$ (solid lines) is compared with numerical simulations
(symbols). In
the simulation we fixed $D_1=D_3=1$ and considered two values of $D_2=0$ and $D_2=1$.
The theoretical predictions of the exponent $\kappa$ in Eq. (\ref{kappa_def}) in these cases
are respectively $\kappa=2$ (for $D_2=0$) and $\kappa=5/2$ (for $D_2=1$).
The agreement between
theoretical predictions and simulations is excellent.}
\label{fig:distt}
\end{figure}

Next, we want to compute the joint distribution $P^{\rm left}(t,z)$ in Eq. (\ref{jpdf_left.1}), by taking the derivative
of Eq. (\ref{polar_late.1}) with respect to $z_1$ and taking the limit $z_1\to 0$. Note that for
$\phi$ and $r^2$ in \eqref{polar_late.1} are both functions of $z_1$. Hence one needs to derive both $\phi$ and $r^2$
with respect to $z_1$ and take the limit $z_1\to 0$. It is however easy to show that
the leading contribution for large $t$ arises from the derivative of the term 
$\sin\left( \frac{\pi (\alpha+\phi)}{\alpha_W}\right)$
with respect to $z_1$ as $z_1\to 0$ in Eq. (\ref{polar_late.1}). We skip the boring alegebraic details here and present only the final result.
We find that in the scaling limit $t\to \infty$, $z_2=z\to \infty$ but with the ratio $z/\sqrt{t}$ fixed, 
the joint distribution $P^{\rm left}(t,z)$, to leading order for large $t$, approaches a scaling form
\begin{eqnarray}
\label{scaling.1}
  P^{\rm left}(t,z) & \approx &
\frac{C}{t^{\theta+3/2}}\, H\left(\frac{z}{\sqrt{4\, \tilde{D}\, t}}\right)\,, 
 \\
   {\rm with} \quad
   \tilde{D} & = & \frac{D_1\,D_2+ D_2\, D_3+ D_3\, D_1}{D_1+D_2}\, .
\nonumber
\end{eqnarray}
Here, $C$ is an overall constant (that depends on initial separations $y_1$ and $y_2$), 
$\theta$ is the persistence exponent given in \eqref{theta_def.1}) and the exact scaling function $H(u)$ 
is given by
\begin{equation}   
H(u)= u^{2\theta-1}\, e^{-u^2} \quad {\rm for}\quad u\ge 0\, .
\label{Hu_def}
\end{equation}
In Fig. \ref{fig:JPDF_tz}, we compare this theoretical prediction
to numerical simulations. Here we used $t_{\max}=100$ and $\Delta t=10^{-4}$
to obtain a good statistics in reasonable time.
Since we rescale the data for different values of $z$ and $t$, the
resulting data points exhibt a considerable scatter, but overall
a very good agreement between numerics and theory is visible.

Having obtained the scaling form \eqref{scaling.1} of the joint distribution $P^{\rm left}(t,z)$, one can
easily derive the tails of the marginal distributions $P^{\rm left}_{t_\ell}(t)$ (by integrating over $z$)
and $P^{\rm left}(z)$ (by integrating over $t$). The former is given by
\begin{eqnarray}
P^{\rm left}_{t_\ell}(t)  & = & \int_0^{\infty} P^{\rm left}(t, z)\, dz \approx \frac{C_1}{t^{\kappa}} 
\quad {\rm as}\quad t\to \infty\, , \nonumber\\
&&  {\rm with}\quad \kappa=\theta+1\, , 
\label{margt.asymp}
\end{eqnarray}
where the amplitude $C_1$ of the power law tail can be easily computed in terms of $C$ in Eq. (\ref{scaling.1})
and the exponent $\theta$ is given in Eq. (\ref{theta_def.1}). Using the expression for $\theta$ in
Eq. (\ref{theta_def.1}), one can express the exponent $\kappa$ as
\begin{equation}
\kappa= 1+ \frac{\pi}{2\, \cos^{-1}\left(\frac{D_2}{\sqrt{(D_1+D_2)(D_2+D_3)}}\right)}\, .
\label{kappa_def}
\end{equation}
This result in Eq. (\ref{margt.asymp}) is clearly consistent with the fact that the survival probability $S(t)\sim t^{-\theta}$ as $t\to \infty$
(see \eqref{surv.2}).
Hence, the total first-passage probability (through the left as well as the right boundary) $F(t)=-dS/dt$ 
defined in Eq. (\ref{sp_fp.1}) decays as $F(t)\sim t^{-\theta-1}$ as $t\to \infty$. Naturally, one would then
expect that the first-passage probability density through the left boundary $F_{\rm left}(t)$, defined in Eq. (\ref{sp_fp.1}),
also decays as a power law with the same exponent $\theta+1$.
Now, the marginal distribution $P^{\rm left}_{t_\ell}(t)$ is just the conditional first-passage time density through
the left boundary and hence is proportional to $F_{\rm left}(t)$, as in Eq. (\ref{marg_tell.1}).
Hence, the power law decay in Eq. (\ref{margt.asymp}) at late times is perfectly consistent with this
observation. In Fig. \ref{fig:distt}, we present numerical simulations to compare with
the theoretical prediction in \eqref{margt.asymp}. 

\begin{figure}
\includegraphics[width=0.47\textwidth]{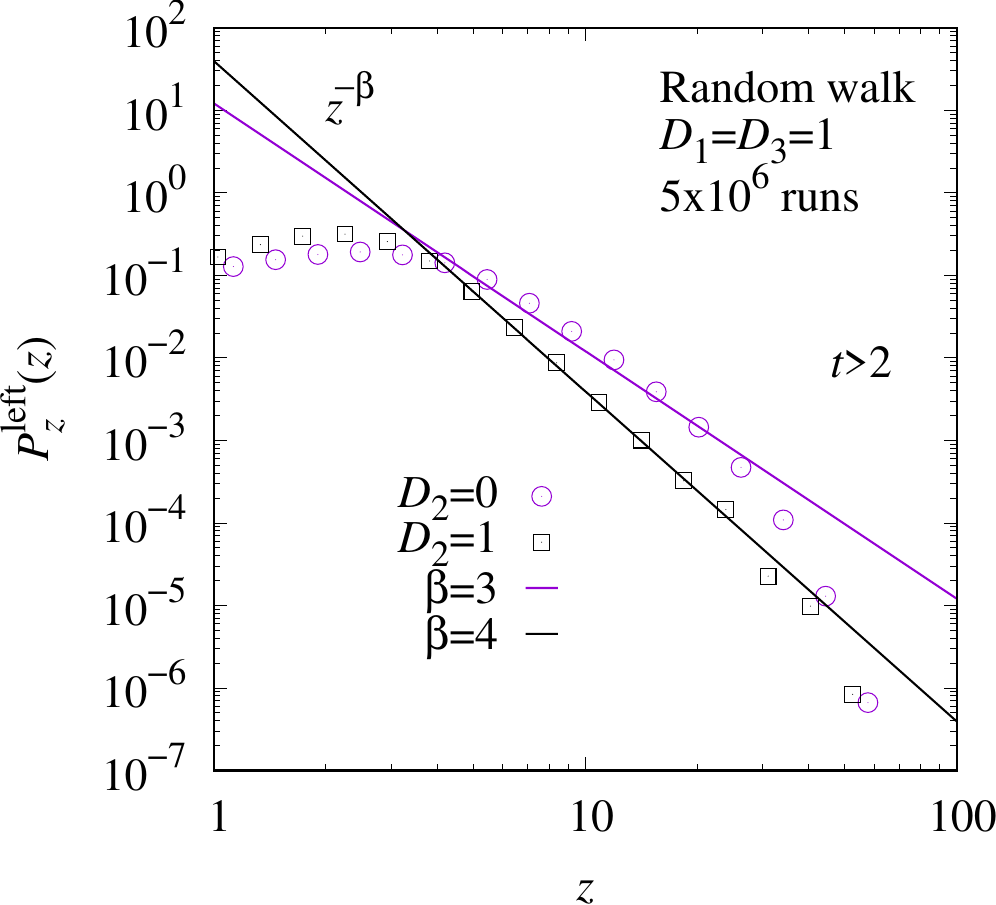}
\caption{Theoretical prediction for the power law decay of the
  marginal distribution $P^{\rm left}_z(z)\sim z^{-\beta}$ in
Eq. (\ref{margz_asymp}) as
a function of $z$ (solid lines) is compared with numerical simulations (symbols). In
the simulation we fixed $D_1=D_3=1$ and considered two values of $D_2=0$ and $D_2=1$ and excluded small splitting times $t\le 2$.
The theoretical predictions of the exponent $\beta$ in Eq. (\ref{beta_def}) in these cases
are respectively $\beta=3$ (for $D_2=0$) and $\beta=4$ (for $D_2=1$). The
agreement between
theoretical predictions and simulations is good for intermediate values
of the distance $z$. }
\label{fig:distz}
\end{figure}

Similarly, the marginal distribution $P^{\rm left}_z(z)$, obtained by integrating \eqref{scaling.1} over $t$,
also decays as a power law for large $z$
\begin{eqnarray}
\label{margz_asymp}
  P^{\rm left}_z(z) & = &
  \int_0^{\infty} P^{\rm left}(t, z)\, dt \approx \frac{C_2}{z^{\beta}}
  \quad {\rm as} \quad z\to \infty\, , \\
  && {\rm where}\quad \beta=2\theta+1\, .
  \nonumber
\end{eqnarray}
The amplitude $C_2$ is again simply related to $C$ in Eq. (\ref{scaling.1}). Using the expression for
$\theta$ in \eqref{theta_def.1}, one can express the exponent $\beta$ as 
\begin{eqnarray}
  \label{beta_def}
  \beta= 2\theta+1 & = &
  1+\frac{\pi}{\cos^{-1}\left(\frac{\gamma}{2}\right)} \\
  & = &
1+\frac{\pi}{\cos^{-1}\left(\frac{D_2}{\sqrt{(D_1+D_2)(D_2+D_3)}}\right)}\,  ,
\nonumber 
\end{eqnarray}
which depends continuously on the diffusion constants $D_1$, $D_2$ and $D_3$.
For example, in the case $D_1=D_2=D_3$, we get $\beta=4$ from \eqref{beta_def}. In contrast, for $D_2=0$,
we get $\beta=3$. We also verified these large $z$ theoretical predictions in our numerical simulations
finding good agreement for intermediate distances, see Fig.~\ref{fig:distz}.
Very large distances are exponentially supressed because
of the finite maximum simulation time $t_{\rm max}$. 

\section{Generalization to fractional Brownian motion}
\label{fBM_section}

We next study a generalization of our three-paricle model where the two outer walkers with positions
$x_1(t)$ and $x_3(t)$ perform
ordinary Brownian motions with diffusion constants $D_1$ and $D_3$ as before, but
the central walker $x_2(t)$ performs a fractional Brownian motion (fBM). 
In the context of the translocation process, this means that the polymerization and depolymerizatiom of 
single monomers at the chain ends on both sides are modeled by a standard random walker (or Brownian motion
in the continuum), whereas the actual translocation through the pore is modeled by an fBM.
As mentioned in the introduction,
the fBM is a Gaussian process with zero mean and a correlator given by~\cite{fBM1968}
\begin{equation}
  \langle x_2(t_1)\,x_2(t_2) \rangle =
  D_2(|t_2|^{2H} + |t_2|^{2H}- |t_2-t_1|^{2H})\,,
\label{eq:correlation}
\end{equation}
with Hurst exponent $0<H<1$.
This leads to a mean-squared displacement (msd) of
\begin{equation}
\langle (x_2(t_0+t)- x_2(t_0))^2 \rangle = 2 D_2 t^{2H}\,. 
\label{eq:msd:H}
\end{equation}
Thus, for $H=1/2$, one recovers the standard Brownian motion.
For $H<1/2$, the long-time behavior of the walk is subdiffusive, i.e., slower as compared
to the $H=1/2$ case, whereas on short time scales $t<1$ it is faster.
For $H>1/2$, the opposite behavior is observed. Furthermore, for $H=1/2$, the process is Markovian,
while for $H\ne 1/2$ it is non-Markovian~\cite{Krug1997,WMR2011}.
Owing to the non-Markovian nature of the process for $H\ne 1/2$, it is hard to compute analytically the
splitting probability as well as the distribution of the time of translocation and that of the chain length 
at the time of translocation.
Hence we compute these observables numerically for $H\ne 1/2$.

To treat the walk numerically, we again consider discrete
time steps $\Delta t$.
The temporal correlations of the position arise from correlations between successive increments
$\Delta x_2(t)=x_2(t+\Delta t)-x_2(t)$. The correlation
$\langle \Delta x_2(t) \, \Delta x_2(t+t') \rangle$ between
two increments evaluates for $D_2=1$
as $|t'+\Delta t|^{2H} + |t'-\Delta t|^{2H} -2|t'|^{2H}$, which
is for $H=1/2$ zero for all values of $t'\ge \Delta t$, as expected, but
non-zero for other values of $H$.

Technically, to generate a realization of a correlated walks,
we first generate 
correlated Gaussian  increments $\tilde \xi(k)$ for unit step size
$\Delta t=1$, i.e., at times $k=1,2,\ldots$, following correlation
Eq.~(\ref{eq:correlation}).
This we achieve efficiently following
the Davies-Harte approach \cite{davies1987}
which is based on Fast Fourier Transform, details can be
found in the literature \cite{HMR2013}. From these unit step
increments $\tilde \xi(k)$ 
we obtain the actual walks, in generalization of Eq. (\ref{eq:step:RW}), as
\begin{equation}
x_2( (k+1) \Delta t ) = x_2(k\Delta t)+ \sqrt{2D} (\Delta t)^H \tilde \xi(k)\,.
\end{equation}

\begin{figure}
\includegraphics[width=0.45\textwidth]{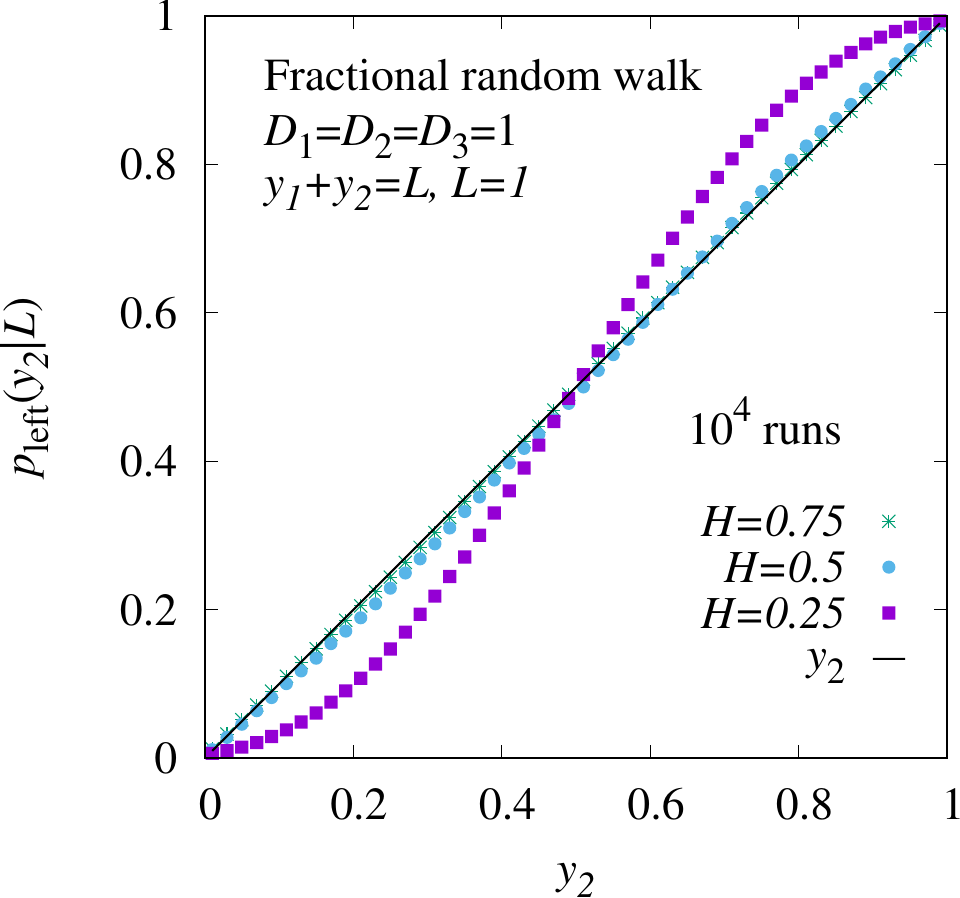}
\includegraphics[width=0.45\textwidth]{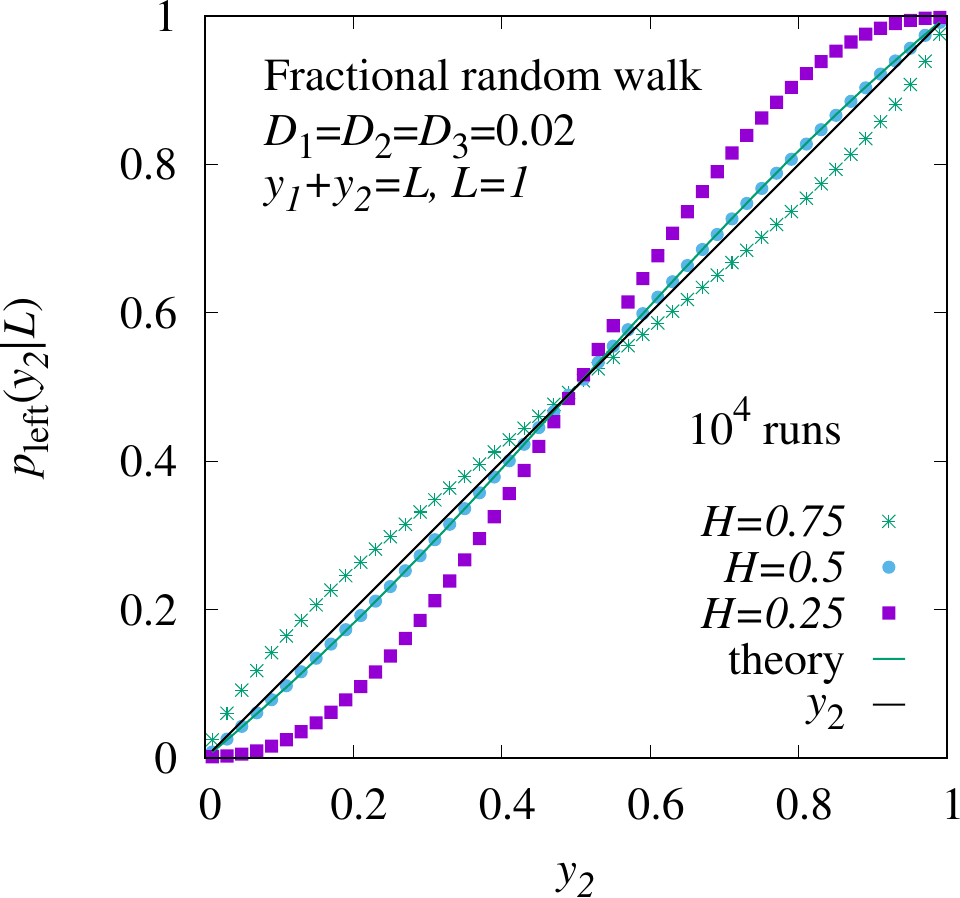}
\caption{The numerically obtained
  splitting probability $p_{\rm left}(y_2|L)$ as
  a function of $y_2$ with $L=1$  for the cases $H=0.25$, $H=0.5$ and $H=0.75$,
  with (top)  $D_1=D_2=D_3=1$, $t_{\max}=10$ and (bottom) $D_1=D_2=D_3=0.2$
  $t_{\max=100}=100$.
\label{fig:hp_H}}
\end{figure}

In Fig.~\ref{fig:hp_H} we show the numerically obtained
hitting probability  $p_{\rm left}(y_2|L)$ as
a function of $y_2$ for $L=1$ and
the cases $H=0.25$, and $H=0.75$, compared to the
uncorrelated case $H=0.5$ as studied above. The upper panel shows the result
for $D_1=D_2=D_3=1$. For $H=0.25$, the deviation
from the linear behavior $p_{\rm left}(y_2|L)=y_2$ is more pronounced
than for the uncorrelated case. Nevertheless, since the diffusion constant
is rather large in comparison with the initial wall distance $L=1$,
the typical splitting times are relatively short.
Consequently, the observed behavior mainly reflects the short-time dynamics..

Thus, we have also considered the case
$D_1=D_2=D_3=0.02$ with still $L=1$.
This changes both the temporal correlations (for $H\ne 1/2$)
and the mean-square displacement.
The particles move more slowly,
so it takes longer for the middle walker to hit a boundary.
Note that alternatively we could have also increased $L$.
Accordingly, we have
used $t_{\max}=100$, with still $\Delta t=10^{-5}$,
so that, in more than $99\%$ of the realizations,
the middle walker hits either the left or right boundary.
As visible in the figure, the behavior for $H=0.75$
changes considerably, while for $H=0.25$ the result is still similar to the
medium-diffusion case.
In general, the latter result is qualitatively  similar to the
behavior that has been found for the case of fractional Brownian motion
 with fixed boundaries
\cite{MRZ2010}.

\begin{figure}
\includegraphics[width=0.45\textwidth]{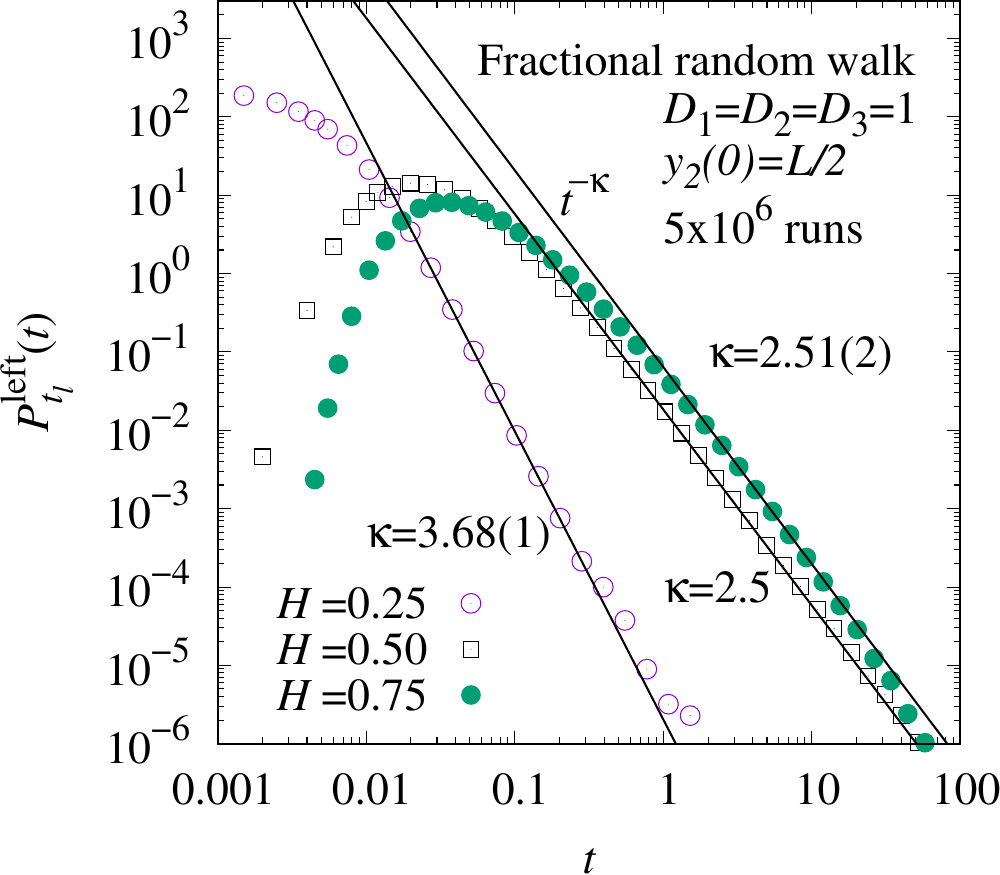}
\includegraphics[width=0.45\textwidth]{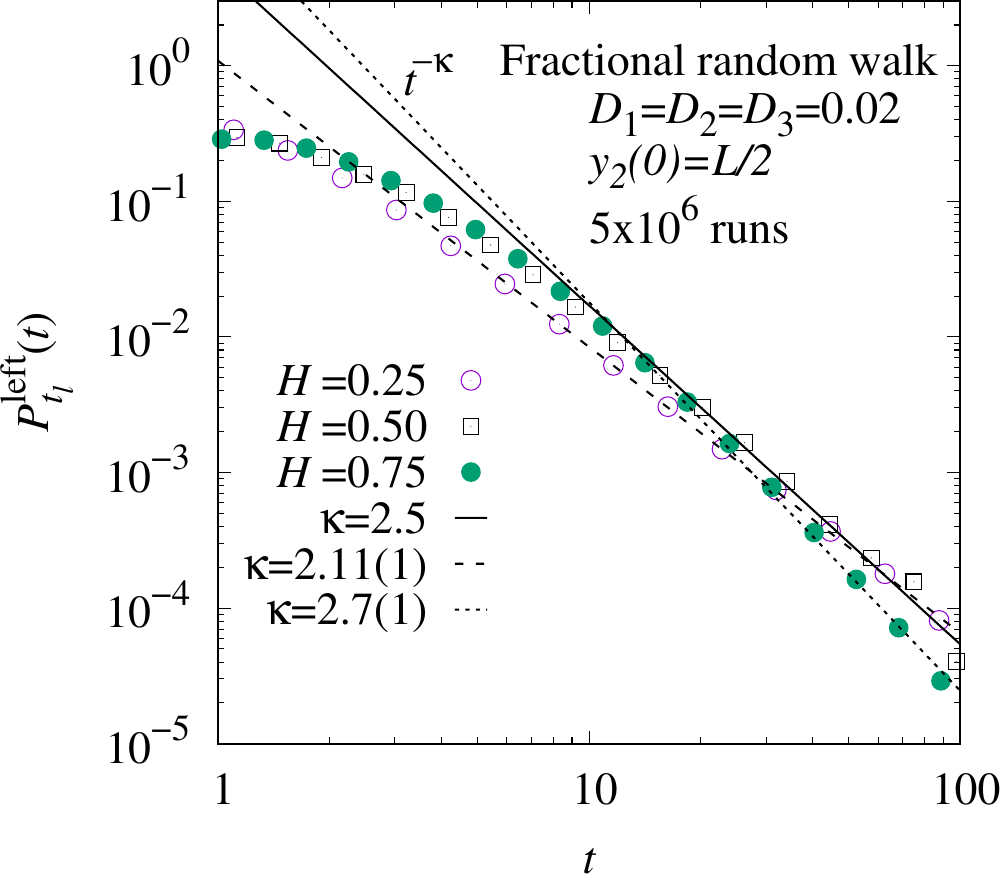}
\caption{Marginal distribution $P^{\rm left}_{t_\ell}(t)$
  obtained from the numerical simulations for $H=0.25$, $H=0.5$
  and $H=0.75$, $L=1$, and for (top) $D_1=D_2=D_3=1$
and (bottom) $D_1=D_2=D_3=0.02$. }
\label{fig:distt:H}
\end{figure}

In Fig.~\ref{fig:distt:H} we show the distribution $P^{\rm left}_{t_\ell}(t)$
of splitting times for the two cases $H=0.25$ and $H=0.75$, and compared them with
the previous result for the uncorrelated case $H=0.5$.
Again, to obtain good statistics within a reasonable computation time, much
larger than for the splitting probability, we
used $t_{\max}=100$ and $\Delta t=10^{-4}$. In the upper plot,
we consider the case $D_1=D_2=D_3=1$.
The
mean-squared displacement given by Eq.~(\ref{eq:msd:H}) for time $t=1$ is always $2D_i=2$
in our model, which is here already of the order of $L=1$.  Since
for small times $t<1$, the msd is 
larger for small values of $H$ as compared to larger values of $H$, the
splitting times $t$ are typically smaller for $H=0.25$ than for $H=0.75$,
as visible in the figure. The data for $H=0.25$
is compatible with power law $\sim t^{-\kappa}$ as for the uncorrelated case,
but with a numerical value $\kappa=3.68(2)$ which is corresponding larger than
the value $\kappa=2.5$. Finally, since the msd for $H=0.75$ is rather small
for $t<1$,  the case $H=0.75$ exhibits larger splitting times. The
actual shape of $P^{\rm left}_{t_\ell}(t)$ is very similar to the
uncorrelated case, a fit to the middle part yields $\kappa=2.51(2)$.

For the slower case $D_1=D_2=D_3=0.02$ the results are shown in the bottom
of Fig.~\ref{fig:distt:H}. Here the splitting times are much larger,
as expected. Interestingly, for the correlated cases $H \neq 1/2$, we
find different exponents $\kappa=2.11(19)$ ($H=0.25$) and $\kappa=2.7(1)$
($H=0.75$) as compared to the case
$D_1=D_2=D_3=1$, while for the uncorrelated case $H=1/2$ we still
observe $\kappa=5/2$, naturally because only the time scale is affected
here, which does not change the distribution of distances upon hitting
the boundaries.
 In general we find the expected long-time
 behavior comparing the results for different values of $H$. Namely,
 for $H=0.25$ the time distributions decreases in the tail slower than
 for $H=0.5$
which itself decreases slower than for $H=0.75$.

\begin{figure}
\includegraphics[width=0.45\textwidth]{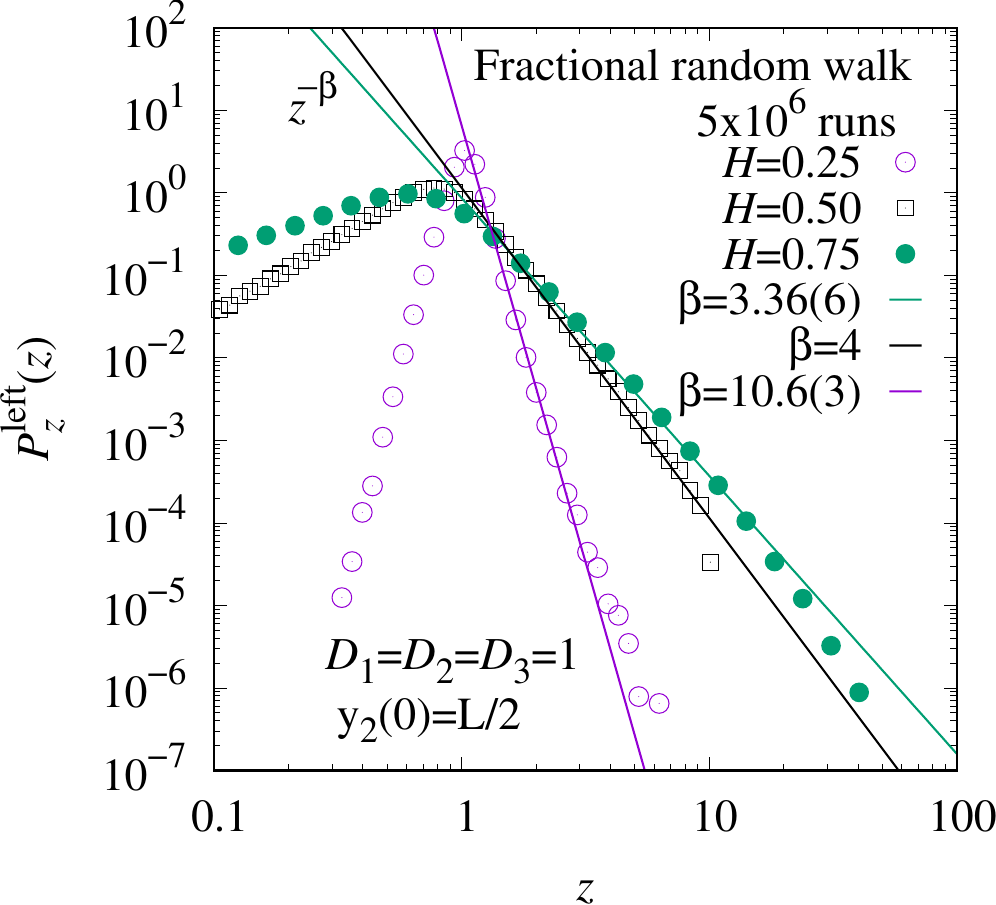}
\includegraphics[width=0.45\textwidth]{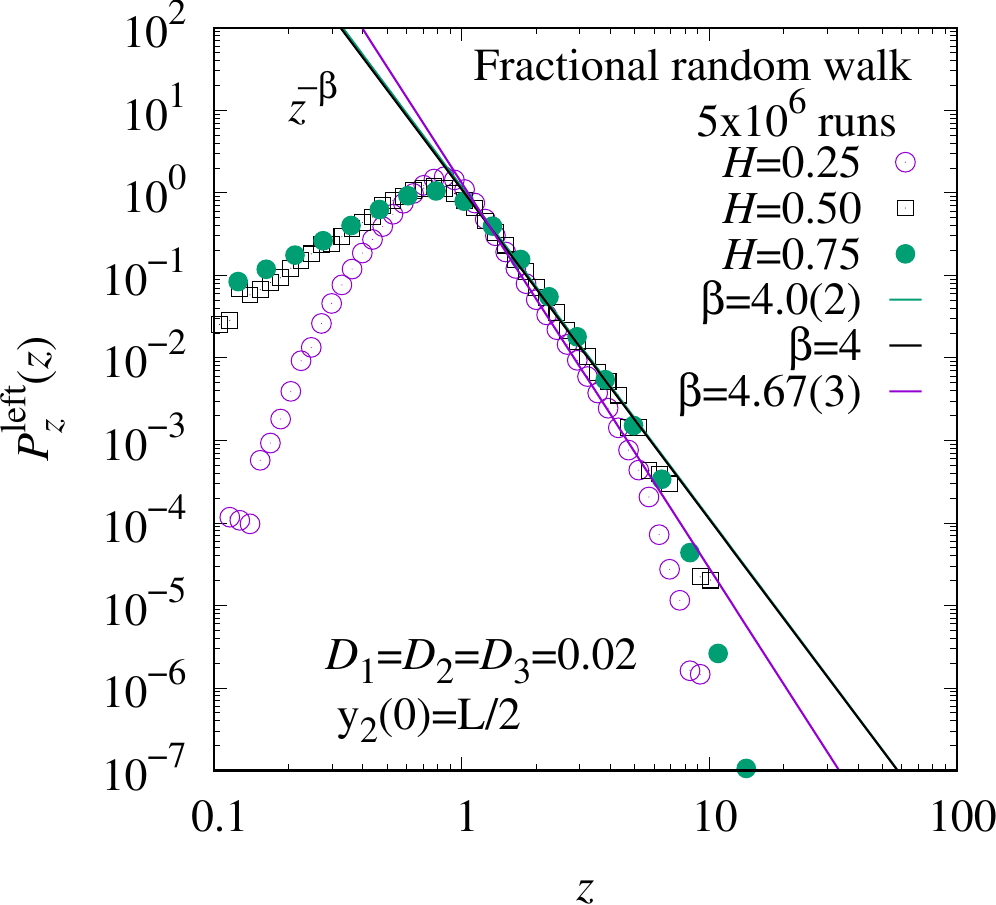}
\caption{Distribution $P^{\rm left}_z(z)$  obtained
  from the numerical simulations for the three cases $H=0.25$, $H=0.5$
  (uncorrelated) and $H=0.75$ for $L=1$. The upper plot is for $D_1=D_2=D_3=1$
  while the lower one for $D_1=D_2=D_3=0.02$.}
\label{fig:distz:H}
\end{figure}

In Fig.~\ref{fig:distz:H} we show the distribution $P^{\rm left}_z(z)$
of the separation $z=y_2$ of the middle walker
to the right boundary at the moment were the middle walker hits the
left boundary, integrated  over all splitting times.
Here, the three cases $H=0.25$,
$H=0.5$ and $H=0.75$ are considered, again for $L=1$ and $x_2(0)=L/2$.

First, we discuss the case $D_1=D_2=D_3=1$,  which is shown on the
upper part of the figure.
We also observe in the right tails
a power law-behavior $\sim z^{-\beta}$ like for the uncorrelated
case at larger times. But since the behavior for $H=0.25$ is
dominated by very small splitting times, see Fig.~\ref{fig:distt:H},
the two boundaries $x_1$ and $x_3$ have not moved much,
so the distance of the two boundaries
is near the initial distance $L=1$, such that the separation $y_2$
between the middle walker and the right boundary is also near $L=1$.
Thus, the distribution falls off very quickly, which is compatible
with a power law with exponent $\beta=10.6(3)$ obtained from
the fit.
The dominance of the distance  $y_2=1$ is also visible
for $H=0.5$ and $H=0.75$ but here larger splitting
times and correspondingly larger distances $y_2$ appear, reflected
by less steep tails. For $H=0.75$, we observe a slightly smaller exponent
than the exact value $\beta=4$ which holds for the uncorrelated case.

On the lower panel of Fig.~\ref{fig:distz:H}, we display the corresponding results
for the slow behavior $D_1=D_2=D_3=0.02$ which leads to larger splitting times,
as discussed. The data for the uncorrelated case does not change as
compared to the case $D_i=1$, except for statistical fluctuations,
since changing all diffusion constants by the same factor only changes
the clock speed for the uncorrelated case. Thus, there is
no change of the exponent $\beta$ in Eq.~(\ref{beta_def}). Now the
distributions of distances
for the three cases are very similar to each other, as can already be
expected from the previously discussed distributions of splitting times.
From power-law fits in Eq.~(\ref{margz_asymp}), we obtain $\beta=4.67(3)$
for $H=0.25$ and $\beta=4.0(2)$ for $H=0.75$, the latter one being
compatible with the
known value $\beta=4$ for the uncorrelated case.

To summarize, when the central particle performs fBM with Hurst exponent $0<H<1$ and 
the two outer particles perform ordinary Brownian motions with 
diffusion constants $D_1$ and $D_3$, the general features of the splitting probability as 
well as the two marginals $P^{\rm left}_{t_\ell}(t)$ and $P^{\rm left}_{z}(z)$ are similar 
to the case when the central particle performs ordinary Brownian motion ($H=1/2$). However, 
the details are different. For example, both the marginal distributions decay as power laws, 
but the associated exponents $\beta$ and $\kappa$ both now depend on $H$, in addition to 
depending on the diffusion constants. 
Thus, although the qualitative behavior remains similar to the Brownian case ($H=1/2$), 
the quantitative results depend on the Hurst exponent $H$.

\section{Conclusion}
\label{summary}

In this work, we studied the translocation process of a polymer chain through a nanopore, 
where the chain length fluctuates due to the stochastic polymerization-depolymerization at 
the chain ends. For the case where the translocation process is modelled by a simple Brownain 
motion, we mapped this process to a problem of three vicious Brownian walkers on a line with 
different diffusion constants $D_1$, $D_2$ and $D_3$. Under this mapping, the translocation 
process terminates when the central walker meets any one of the two outer walkers. We 
computed exactly three important observables: (i) the splitting probability, i.e., the 
probability that the central walker meets the walker on its left (right) before that on the 
right (left). In the language of translocation, this splitting probability is the same as 
the probability that the chain translocates to the right (left) of the pore. (ii) the 
probability distribution of the time at which the translocation process terminates and (iii) 
the probability distribution of the length of the polymer chain at the time when the 
translocation process terminates. We showed that the splitting probability is rather 
nontrivial in the case when the two outer particles diffuse, compared to the classical 
well-known case when the two outer particles are fixed (the latter corresponds to the case 
where there is no polymerization-depolymerization at the chain ends and hence the chain 
length is fixed). We also showed that the distribution of the translocation time and that of 
the chain length at the completion time of translocation both have power-law tails with 
exponents that depend cotinuously on the three diffusion constants $D_1$, $D_2$ and $D_3$. 
All analytical predictions were validated by extensive numerical simulations, showing 
excellent agreement over the entire parameter range investigated.

Finally, we extended the model by replacing the Brownian motion of the pore with fractional 
Brownian motion while keeping the polymer ends diffusive. Owing to the non-Markovian nature 
of fractional Brownian motion, an analytical treatment is considerably more challenging, and 
we therefore relied on numerical simulations. We found that the qualitative features of the 
Brownian case remain robust, while the splitting probabilities and the asymptotic exponents 
depend continuously on the Hurst exponent. In particular, the persistence exponent and the 
exponent governing the distribution of the final polymer length vary systematically with the 
degree of temporal correlations introduced by the fractional Brownian motion.

The present work provides a unified framework for studying first-passage phenomena in 
systems with fluctuating boundaries and establishes several exact results for Brownian 
dynamics together with numerical results for their fractional Brownian generalization. 
Beyond the polymer translocation problem that motivated this study, the methods developed 
here may be applicable to a broader class of first-passage problems involving moving 
boundaries or multiple interacting stochastic processes. An interesting direction for future 
work would be to develop an analytical treatment of the fractional Brownian case and to 
investigate more realistic models that incorporate interactions, external driving forces, or 
additional internal degrees of freedom of the polymer.

\begin{acknowledgments}

SNM acknowledges support from ANR Grant No. ANR- 23- CE30-0020-01 EDIPS and the Alexander 
von Humboldt foundation for the Gay Lussac-Humboldt prize that allowed a visit to the 
Physics department at Oldenburg University, Germany where part of this work was performed. 
The simulations were performed at the HPC cluster ROSA, located at the University of 
Oldenburg (Germany) and funded by the DFG through its Major Research Instrumentation Program 
(INST 184/225-1 FUGG) and the Ministry of Science and Culture (MWK) of the Lower Saxony 
State.

\end{acknowledgments}

\end{document}